\documentclass[5p,authoryear]{elsarticle}
\usepackage{mathtools}
\usepackage{amsmath}
\usepackage{graphicx}
\usepackage{times}
\usepackage{amssymb}
\usepackage{multirow}
\usepackage{algorithm}
\usepackage{algorithmicx}
\usepackage{algpseudocode}
\usepackage{color}
\usepackage{textcomp}
\usepackage{eqparbox}
\usepackage{hyperref}
\usepackage{wrapfig}
\usepackage{subfig}
\usepackage{tabularx}
\usepackage[explicit]{titlesec}

\titleformat{\section}{\normalfont}{\textbf{\thesection}}{1em}{\MakeUppercase{\textbf{#1}}}
\titlespacing*{\section}{0em}{2em}{0.8em}
\titleformat{\subsection}{\normalfont}{\textbf{\thesubsection}}{1em}{\textbf{#1}}
\titlespacing*{\subsection}{0em}{2em}{0.8em}
\titleformat{\subsubsection}{\normalfont}{\textit{\thesubsubsection}}{1em}{\textit{#1}}
\titlespacing*{\subsubsection}{0em}{2em}{0.8em}

\newcommand{\hompc}{\,h\,{\rm Mpc}^{-1}}
\newcommand{\mpcoh}{\,h^{-1}\,{\rm Mpc}}
\newcommand{\gpcoh}{\,h^{-1}\,{\rm Gpc}}

\begin{document}

% --- title --- %
\title{L-PICOLA: A parallel code for fast dark matter simulation}

\author[icg]{Cullan Howlett\corref{email1}}
\author[ucl,icg]{Marc Manera}
\author[icg]{Will J. Percival}
\address[icg]{Institute of Cosmology \& Gravitation, Dennis Sciama Building, University of Portsmouth, Portsmouth, PO1 3FX, UK}
\address[ucl]{University College London, Gower Street, London WC1E 6BT, UK.}
\cortext[email1]{cullan.howlett@port.ac.uk}
\date{outline} 

\begin{abstract}
Robust measurements based on current large-scale structure surveys require precise knowledge of statistical and systematic errors. This can be obtained from large numbers of realistic mock galaxy catalogues that mimic the observed distribution of galaxies within the survey volume. To this end we present a fast, distributed-memory, planar-parallel code, {\sc l-picola}, which can be used to generate and evolve a set of initial conditions into a dark matter field much faster than a full non-linear N-Body simulation. Additionally, {\sc l-picola} has the ability to include primordial non-Gaussianity in the simulation and simulate the past lightcone at run-time, with optional replication of the simulation volume. Through comparisons to fully non-linear N-Body simulations we find that our code can reproduce the $z=0$ power spectrum and reduced bispectrum of dark matter to within $2\%$ and $5\%$ respectively on all scales of interest to measurements of Baryon Acoustic Oscillations and Redshift Space Distortions, but 3 orders of magnitude faster. The accuracy, speed and scalability of this code, alongside the additional features we have implemented, make it extremely useful for both current and next generation large-scale structure surveys. {\sc l-picola} is publicly available at \url{https://cullanhowlett.github.io/l-picola}.

\end{abstract} 


\maketitle

\section{Introduction}
Analysis of the large scale structure of the universe allows us to probe the universe throughout its expansion history and provides the most robust route to measuring the late-time evolution of the universe. Over the last decade, large sky-area galaxy surveys such as the Sloan Digital Sky Survey (SDSS; \citealt{York2000,Eisenstein2011}), 2dF Galaxy Redshift Survey (2dFGRS; \citealt{Colless2001,Colless2003}), 6dF Galaxy Redshift Survey (6dFGRS, \citealt{Jones2004, Jones2009}) and WiggleZ survey \citep{Drinkwater2010} have allowed us to probe this large scale structure and have provided us with a wealth of cosmological information. In particular measurements of the Baryon Acoustic Oscillation scale (BAO; \citealt{Seo2003,Eisenstein2005}) provide us with a standard ruler, allowing us to measure the accelerated expansion of the universe, whilst Redshift Space Distortions (RSD; \citealt{Kaiser1987}) provide a direct probe of the growth of structure and the fidelity of General Relativity. These probes have become more and more accurate in recent years, with \cite{Anderson2014} providing a 1\% measurement of the distance scale to $z=0.57$, the most precise from a galaxy survey to date. However, these measurements, and their errors, require intimate knowledge of the statistical and systematic distributions from which they are drawn. This need will only be exacerbated as future surveys, such as the Large Sky Synoptic Telescope (LSST; \citealt{Ivezic2008}) and Euclid \citep{Laureijs2011}, strive for even greater precision.

Under the assumption of Gaussian-distributed errors, the statistical errors inherent in large scale clustering measurements are encapsulated by the covariance matrix. Although this can be calculated analytically in the linear regime \citep{Tegmark1997}, the non-linear galaxy covariance matrix is a complex function of non-linear shot-noise, galaxy evolution and the unknown relationship between the galaxies and the underlying dark matter. In any real application this is further compounded by the effect of RSD. As such, a much more common solution is to use a set of detailed galaxy simulations, otherwise known as mock catalogues (mocks), to either fully estimate the covariance matrix or as the basis for an empirically motivated analytic fit \citep{Xu2012}.

Ideally these simulations would take the form of fully realised N-Body simulations, with accurate small scale clustering, covering the whole volume of the galaxy survey. However, for current surveys, recent studies \citep{Dodelson2013, Taylor2013, Percival2014} show that we require 1000 mocks to obtain an accurate numerical estimate of the covariance matrix with sub-dominant errors compared to the statistical errors themselves. Higher precision measurements in the future may require many more. Instead there have been many studies looking at fast methods of producing simulations that enable us to produce mocks hundreds of times faster than an Tree-PM N-Body simulation, at the cost of reduced small scale clustering accuracy.

Past measurements of the large scale structure have used lognormal models to generate realizations of the galaxy overdensity field and estimate the covariance matrix \citep{Coles1991,Cole2005}. However this approach does not accurately capture the non-Gaussian behaviour of gravitational collapse. Recently, \cite{Manera2013} and \cite{Manera2015} have used an implementation of the more accurate PTHALOS method \citep{Scoccimarro2002} to generate mock catalogues for the Data Releases 9 and 10 of the Baryon Oscillation Spectroscopic Survey (BOSS; \citealt{Ahn2012,Ahn2014} \\ \citealt{Dawson2013}). 

In addition to this there is a wealth of alternative methods such as PINOCCHIO \citep{Monaco2002,Monaco2013}, Quick Particle Mesh Simulations (QPM; \citealt{White2014}), Augmented lagrangian Perturbation Theory (ALPT;  \citealt{Kitaura2013}), Effective Zel\textquotesingle dovich approximation mocks (EZmocks; \citealt{Chuang2015}) and the Comoving Lagrangian Acceleration method (COLA; \citealt{Tassev2013}, \citealt{Tassev2015}) which can all be used to produce mock catalogues comparable in accuracy to, if not better than, 2LPT and with similar speed. In this paper we present a stand-alone parallel implementation of the latter of these, with emphasis on maximising speed, memory conservation and ease of use. This code, which we dub {\sc l-picola}, combines a range of features that will be of increasing interest for the next generation of galaxy surveys, including the ability to produce lightcone simulations, replicate the simulation at runtime and include primordial non-Gaussianity based on a generic input bispectrum, as per \cite{Scoccimarro2012}. On top of this the COLA method itself is able to reproduce the dark matter field with much greater accuracy on small, non-linear scales than the PTHALOS method, at only a moderate increase in computational cost. 

As such, we expect our implementation to be suitable for both current and future surveys, being able to both capture non-linear evolution with a precision necessary to reach the required covariance matrix accuracy for these surveys and scalable up to very large numbers of particles and volumes. In fact, we have already used this code to measure the BAO and RSD signals from a subset of luminous red galaxies drawn from the Sloan Digital Sky Survey Data Release 7 Main Galaxy Sample \citep{Ross2015, Howlett2015}. Additionally, Manera et al., (in prep.) describe an application of {\sc l-picola} to the Dark Energy Survey (DES; \citealt{DES2005}), making use of the fast lightcone algorithm that we will discuss in this paper.

It should be noted that in previous studies using this code, we named the code {\sc picola}. However, very recently, \cite{Tassev2015} present an extension to the COLA method that allows one to decouple the short and long range gravitational forces \textit{spatially} in addition to temporally. This allows the user to calculate the non-linear displacements for only a small subsection of the full simulation volume and still recover reasonable accuracy across the whole simulation. In this work they also present a code {\sc pyCOLA}, a shared-memory Python implementation of the extended COLA method. To avoid confusion in the names of these codes, and highlight the additional features we have implemented since the first application of our code, we have renamed it {\sc l-picola}. 

This paper is structured as follows: In Section 2 we provide a brief description of the theory behind the 2LPT and COLA methods. Section 3 introduces {\sc l-picola}, with Section 4 detailing the steps we have taken to parallelise the COLA method for a distributed-memory machine. Sections 5 and 6 detail the additional features we have included in {\sc l-picola}, beyond the standard snapshot simulations. In particular, in Section 6 we validate the need for lightcone simulations and perform a rigorous test of our implementation. In Sections 7 we compare the accuracy of {\sc l-picola} to 2LPT and a full N-Body simulation. In this section we also test the effect on the clustering accuracy of several of the free parameters that can be used to speed up the convergence of the COLA method. In Section 8 we compare the speed of {\sc l-picola} with the 2LPT and N-Body simulations, and look at the scaling of different segments of {\sc l-picola} itself. Finally in Section 9 we conclude and discuss further improvements that can be made to the code. Included in the Appendix are details of the memory footprint of {\sc l-picola}.

Unless otherwise stated, we assume a fiducial cosmology given by $\Omega_m = 0.317$, $\Omega_b=0.049$, $h=0.67$, $\sigma_{8}=0.83$, and $n_s=1.0$. Also, unless otherwise stated, all simulations presented use a number of mesh cells equal to the number of particles, the COLA method with modified COLA timestepping, $nLPT=-2.5$ and 10 linearly spaced timesteps. These {\sc l-picola}-specific parameters are stated here for completeness but are explained within this paper.

\section{The COmoving Lagrangian Acceleration (COLA) method}

In the following section we describe the COLA method for evolving a system of dark matter particles under gravity, as first introduced by \cite{Tassev2013}. We begin with a summary of the theoretical underpinnings of the algorithm, including a brief overview of second order lagrangian perturbation theory (2LPT), before moving onto the algorithmic implementation.

\subsection{2LPT}
As described in \cite{Scoccimarro1998} (see also \citealt{Moutarde1991} and \citealt{Bouchet1995}), cold dark matter particles evolving over cosmological time in an expanding universe follow the equation of motion (EOM)
\begin{equation}
\frac{d^{2}\boldsymbol{\Psi}}{d\tau^{2}} + \mathcal{H}(\tau)\frac{d\boldsymbol{\Psi}}{d\tau} + \nabla \Phi = 0,
\label{eom}
\end{equation}
where $\Phi$ is the gravitational potential, $\mathcal{H}(\tau) = \frac{dlna}{d\tau}$ is the conformal Hubble parameter and $a$ is the scale factor. $\boldsymbol{\Psi}$ is the displacement vector of the particle and relates the particle's Eulerian position $\boldsymbol{x}(\tau)$ to its initial, Lagrangian position, $\boldsymbol{q}$, via
\begin{equation}
\label{lagrangian}
\boldsymbol{x}(\tau) = \boldsymbol{q} + \Psi(\boldsymbol{q},\tau).
\end{equation}
By taking the divergence of the equation of motion and using the Poisson equation, we find
\begin{equation}
\nabla_{x} \cdot \left( \frac{d^{2}\boldsymbol{\Psi}}{d\tau^{2}} + \mathcal{H}(\tau)\frac{d\boldsymbol{\Psi}}{d\tau} \right) = - \frac{3}{2} \Omega_{m,0} \mathcal{H}(\tau) \delta(\tau).
\label{eom2}
\end{equation}
Here $\Omega_{m,0}$ is the matter density at $\tau=0$, whilst $\delta(\tau)$ is the local overdensity.
Lagrangian perturbation theory seeks to solve this equation by perturbatively expanding the displacement vector,
\begin{equation}
\boldsymbol{\Psi} = \boldsymbol{\Psi}^{(1)} + \boldsymbol{\Psi}^{(2)} + \ldots,
\end{equation}
If we then apply the continuity equation, $\rho(\boldsymbol{x},t)d^{3}x = \rho(\boldsymbol{q},0)d^{3}q$, which states that a mass element $d^{3}q$ centred at $\boldsymbol{q}$ at time zero becomes a mass element $d^{3}x$, centred at $\boldsymbol{x}$, at time $t$, we find that, to first order
\begin{equation}
\nabla_{q} \cdot \boldsymbol{\Psi}^{(1)} = -D_{1}(\tau) \delta_{L}(\boldsymbol{q}).
\label{za}
\end{equation}
This is the well known Zel\textquotesingle dovich appoximation (ZA; \citealt{Zeldovich1970}). Here $D_{1}(\tau)$ is the linear growth factor, $\delta_{L}(q)$ is the linear overdensity field and we have rewritten the divergence as a function of $\boldsymbol{q}$ by using the fact that they are related via the Jacobian of the transformation from $\boldsymbol{x}$ to $\boldsymbol{q}$, i.e., $\nabla x_{i} = (\delta_{ij} + \partial \boldsymbol{\Psi}_{i}/\partial \boldsymbol{q}_{j})^{-1} \nabla q_{j}$.
Solving to second order introduces corrections to the first order displacement of the form
\begin{equation}
\nabla_{q} \cdot \boldsymbol{\Psi}\,^{(2)}= \frac{1}{2}D_{2}(\tau)\sum_{i\neq j}\left(\boldsymbol{\Psi}^{(1)}_{i,i}\boldsymbol{\Psi}^{(1)}_{j,j} - \boldsymbol{\Psi}^{(1)}_{i,j}\boldsymbol{\Psi}^{(1)}_{j,i} \right),
\label{2lpt}
\end{equation}
where, for brevity, we have defined $\boldsymbol{\Psi}_{i,j}=\partial \boldsymbol{\Psi}_{i}/\partial \boldsymbol{q}_{j}$.
\cite{Bouchet1995} provide a good approximation for $D_{2}(\tau)$, the second order growth factor, for a flat universe with non-zero cosmological constant
\begin{equation}
D_{2}(\tau) \approx -\frac{3}{7}D^{2}_{1}(\tau)\Omega_{m}(\tau)^{-1/143}. 
\label{eq:2LPTgrowth}
\end{equation}
For further computational ease, we can define two Langragian potentials, $\boldsymbol{\Psi}^{(i)}=\nabla_{q}\phi^{(i)}$, such that Eq. (\ref{lagrangian}) becomes
\begin{equation}
\boldsymbol{x}(\tau) = \boldsymbol{q} - D_{1}\nabla_{q}\phi^{(1)} + D_{2}\nabla\phi^{(2)},
\end{equation}
and the Lagrangian potentials are obtained by solving the corresponding pair of Poisson equations derived from Eq. (\ref{za}) and Eq. (\ref{2lpt}) respectively,
\begin{equation}
\nabla_{q}^{2}\phi^{(1)} = \delta_{L}(\boldsymbol{q}).
\end{equation}
\begin{equation}
\nabla_{q}^{2}\phi^{(2)} =\sum_{i>j}\left(\phi_{i,i}^{(1)}\phi_{j,j}^{(1)} - (\phi_{i,j}^{(1)})^{2}\right).
\end{equation}

\subsection{COLA}
The COLA method \citep{Tassev2013} provides a much more accurate solution to Eq.(\ref{eom}) than 2LPT, at only a moderate ($\sim 3\times$) reduction in speed. It does this by utilising the first and second-order lagrangian displacements, which provide an exact solution at large, quasi-linear scales, and solving for the resultant, non-linear component. By switching to a frame of reference comoving with the particles in Lagrangian space, we can split the dark matter equation of state as follows,
\begin{equation}
T[\boldsymbol{\Psi}_{res}]+ T[{D}_{1}]\boldsymbol{\Psi}_{1} + T[{D}_{2}]\boldsymbol{\Psi}_{2} + \nabla \Phi = 0,
\label{eq:cola}
\end{equation}
where,
\begin{equation}
T[X] = \frac{d^{2}X}{d\tau^{2}} + \mathcal{H}\frac{dX}{d\tau}.
\end{equation}
$\boldsymbol{\Psi}_{res}$ is the remaining displacement when we subtract the quasi-linear 2LPT displacements from the full, non-linear displacement each particle should actually feel. 

The reason this method is so useful is because we only need to calculate the Lagrangian displacements once, at redshift $z=0$, and scale them by the appropriate derivatives of the growth factor. In fact, as we will see in later sections, it is common practice in many N-Body simulations to use 2LPT to generate the initial positions of the particles at a suitably high redshift, where the results are exact.

In {\sc l-picola}, Eq. (\ref{eq:cola}) is solved as a whole (as opposed to evaluating $\Psi_{res}$ individually) by discretising the operator `T' using the Kick-Drift-Kick algorithm \citep{Quinn1997}, such that at each iteration the velocity and position of each particle is updated based on the gravitational potential $\Phi$ \textit{and} the stored 2LPT displacements. The well-known Particle-Mesh algorithm, with forward (FFT) and inverse (IFFT) Fourier transforms, is used to evaluate the gradient of $\Phi$ using the particle density. I.e,
\begin{equation}
\nabla{\Phi} = \text{IFFT}\left[\frac{3}{2}\frac{\Omega_{m,0}\boldsymbol{k}}{ak^2}\times\text{FFT}[\rho(\boldsymbol{x})-1]\right]
\end{equation}
The following sections detail the Kick-Drift-Kick method and the Particle-Mesh algorithm used in {\sc l-picola} and how these are easily modified to solve Eq. (\ref{eq:cola}) as opposed to the standard dark matter equation of motion.

\subsection{Timestepping} \label{sec:timestepping}

Eq.~\ref{eq:cola} is discretised using the Kick-Drift-Kick/Leapfrog \\ method \citep{Quinn1997}. The modified, COLA dark matter EOM is solved iteratively, and at each iteration the particle velocities and positions are updated based on the gravitational potential felt by each particle. Particle velocities are calculated from the displacements and updated to the nearest half-integer timestep. The particle positions are then updated to the nearest integer timestep using the previous velocity. In this way the particle velocities and positions are never (except at the beginning and end) calculated for the same point in time but rather `leapfrog' over each other with the next iteration of the velocity dependent on the position from the previous iteration and so on. 

In the standard, non-COLA method, the dark matter EOM can be solved via.  
\begin{align}
&\boldsymbol{v}_{i+1/2} = \boldsymbol{v}_{i-1/2} - \nabla\phi\Delta a_{1}, \label{eq:vcola} \\
&\boldsymbol{r}_{i+1} = \boldsymbol{r}_{i} + \boldsymbol{v}_{i+1/2}\Delta a_{2} \label{eq:rstandard}
\end{align}
$\Delta a_{i}$ encapsulates the time interval and appropriate numerical factors required to convert the displacement to a velocity and the velocity to a position. \cite{Quinn1997} evaluate these as
\begin{align}
\Delta a_{1} &= \frac{H_{0}}{a_{i}}\int_{a_{i-1/2}}^{a_{i+1/2}} \frac{da}{a^{2}H(a)}, \notag \\
\Delta a_{2} &= H_{0}\int_{a_{i}}^{a_{i+1}}\frac{da}{a^{3}H(a)}.
\end{align}

The equations for updating the particle positions and velocities can be modified to solve the COLA EOM in the following way
\begin{align}
&\boldsymbol{v}_{i+1/2} = \boldsymbol{v}_{i-1/2} - T[\boldsymbol{\Psi}_{res}]\Delta a_{1}, \label{eq:vcola} \\
&\boldsymbol{r}_{i+1} = \boldsymbol{r}_{i} + \boldsymbol{v}_{i+1/2}\Delta a_{2} + \Delta D_{1}\boldsymbol{\Psi}_{1} + \Delta D_{2}\boldsymbol{\Psi}_{2} \label{eq:rcola}
\end{align}
Here $\Delta D = D_{i+1} - D_{i}$ denotes the change in the first and second order growth factors over the timestep. The modified Kick-Drift-Kick equations are derived under the condition that, for Eq. (\ref{eq:cola}) to be valid, the displacements felt by each particle must be computed in the 2LPT reference frame. In other words, the acceleration each particle feels due to the the gravitational potential must be modified, and the 2LPT contribution to the acceleration removed. The new gravitational potential is then, by design, $T[\Psi_{res}]$. The exact procedure used to calculate $\nabla\Phi$ is not important and as such any code that updates the particle velocities and positions iteratively based on the gravitational potential, i.e., a Tree-PM code, can be modified in the above way to include the COLA mechanism.

An important point of note in enforcing the change of reference frame is that particle velocities at the beginning of the simulation, after creation of the 2LPT initial conditions but before iterating, must be identically 0. At this point the velocity a particle has is exactly equal to the velocity of the reference frame we are moving too. However when the particles are output at the end of the simulation we want the particle velocities in \textit{Eulerian} coordinates. This means that the initial particle velocities must be removed and stored at the beginning of the timestepping and then added back on at the end of the simulation.

When implementing the modified COLA timestepping, the time intervals, $\Delta a_{i}$, for each timestep do not get explicitly changed and as such can remain the same as the those presented in \cite{Quinn1997}. However, \cite{Tassev2013} present a second, COLA specific, formulation which gives faster convergence, hence allowing us to recover our evolved dark matter field to greater accuracy in fewer time steps. In their method,
\begin{align}
\Delta a_{1} &= \frac{H_{0}}{nLPT}\frac{a_{i+1/2}^{nLPT}-a_{i-1/2}^{nLPT}}{a_{i}^{nLPT-1}}, \notag \\
\Delta a_{2} &= \frac{H_{0}}{a_{i+1/2}^{nLPT}}\int_{a_{i}}^{a_{i+1}}\frac{a^{nLPT-3}}{H(a)} da.
\label{eq:modifiedt}
\end{align}
where they find the best results using a value $nLPT=2.5$. As the choice of $\Delta a_{i}$ is somewhat arbitrary {\sc l-picola} retains both methods as options. This choice (and $nLPT$) should be treated formally as an extra degree of freedom in the code. In fact we find that the value of $nLPT$ used in the code can affect the final shape of the power spectrum recovered from {\sc COLA} due to the way different growing modes are emphasised by different values. This is pointed out in \cite{Tassev2013} and means that for a given set of simulation parameters one would ideally experiment to find the type of timestepping that recovers the required clustering in the fewest timesteps possible. This is demonstrated further in Section \ref{sec:accuracy}. 

For the timestepping method presented here and adopted in {\sc l-picola}, the only remaining piece of the puzzle is the calculation of $T[\boldsymbol{\Psi}_{res}] = -T[{D}_{1}]\boldsymbol{\Psi}_{1} - T[{D}_{2}]\boldsymbol{\Psi}_{2} - \nabla \Phi$. As the ZA and 2LPT displacements have been stored we only need a method of evaluating $\nabla \Phi$.  In {\sc l-picola} this is done using the Particle-Mesh algorithm, though could be done using a method such as the Tree-PM algorithm. The evaluation of $T[{D}_{1}]$ and $T[{D}_{2}]$ can be performed numerically for a given cosmological model very easily, although a suitable approximation for $D_{2}$ must be adopted. For flat cosmologies one could use Eq (\ref{eq:2LPTgrowth}), however in {\sc l-picola} we adopt the expression of \cite{Matsubara1995} which is also accurate for non-flat cosmologies.

\subsection{Particle-Mesh algorithm}

Here we provide a brief overview of the Particle-Mesh (PM) algorithm (see \cite{Hockney1988} for a good review of this method). Our implementation is based on the publicly available {\sc pmcode}\footnote{\url{http://astro.nmsu.edu/~aklypin/PM/pmcode/}} \citep{Klypin1997}, and as such we refer the reader to the associated documentation for full details on the set of equations we solve to get the displacement. 

In the PM method we place a mesh over our dark matter particles and solve for the gravitational forces at each mesh point. We then interpolate to find the force at the position of each particle and use this to calculate the gravitational potential each particle feels. This gravitational potential is then related to the additional velocity, and resultant displacement, for each particle as per the This is performed iteratively over a series of small timesteps. For $N_{m}^{3}$ mesh points and $N^{3}$ particles, this means that at each iteration we only need to perform $N_{m}^{3}$ force calculations, which is much faster than a direct summation of the contribution to the gravitational force from each individual particle (at least for all practical applications, where $N \approx N_{m}$). 

At each iteration we perform the following steps to calculate the displacement:
\begin{enumerate}
\item{Use the Cloud-in-Cell linear interpolation method to assign the particles to the mesh, thereby calculating the mass density, $\rho(\boldsymbol{x})$, at each mesh point.}
\item{Use a Fast Fourier Transform (FFT) to Fourier transform the density and solve the comoving Poisson equation in Fourier space\footnote{In all cases we use the FFTW-3 Discrete Fourier Transform routines to compute our Fourier transforms. This library is freely available from \url{http://www.fftw.org/}}.}
\begin{equation}
k^2 \phi(\boldsymbol{k}) =  \frac{3}{2}\frac{\Omega_{m,0}}{a}(\rho(\boldsymbol{k})-1)
\end{equation}
\item{Use the gravitational potential and an inverse-FFT to generate the force in each direction in real-space. Here we also deconvolve the Cloud-in-Cell window function.}
\begin{equation}
F(\boldsymbol{k}) = \boldsymbol{k}\phi(\boldsymbol{k})
\end{equation}
\item{Calculate the acceleration each particle receives in each direction, again using the Cloud-in-Cell interpolation method to interpolate from the mesh points.}
\end{enumerate}

\section{A Lightcone-enabled Parallel Implementation of COLA (L-PICOLA)}

As suggested by the name, {\sc l-picola} is a parallel implementation of the COLA method described in the previous section. We have designed the code to be `stand alone', in the sense that we can generate a dark matter realisation based solely on a small number of user defined parameters. This includes preparing the initial linear dark matter power spectrum, generating an initial particle distribution with k-space distribution that matches this power spectrum, and evolving the dark matter field over a series on user specified timesteps until some final redshift is reached. At any point in the simulation the particle position and velocities can be output, allowing us to capture the dark matter field across a variety of epochs in a single simulation. 

In order to make {\sc l-picola} as useful as possible we have also implemented several options that modify how {\sc l-picola} is actually built at compile time. On top of allowing variations in output format and memory/speed balancing we also allow the user to create (and then evolve) initial particle distributions containing primordial non-Gaussianity. Another significant improvement, and one that will be extremely important for future large scale structure surveys, is the option to generate a lightcone simulation, which contains variable clustering as a function of distance from the observer, as opposed to a snapshot simulation at one fixed redshift. Although lightcone simulations can be reconstructed from a series of snapshots \citep{Fosalba2013, Merson2013}, {\sc l-picola} can produce lightcone simulations `on-the-fly' in a short enough time to be suitable for generating significant numbers of mock galaxy catalogues. These additions will be detailed and tested in later sections.

\begin{figure*}
\centering
\includegraphics[width=0.8\textwidth]{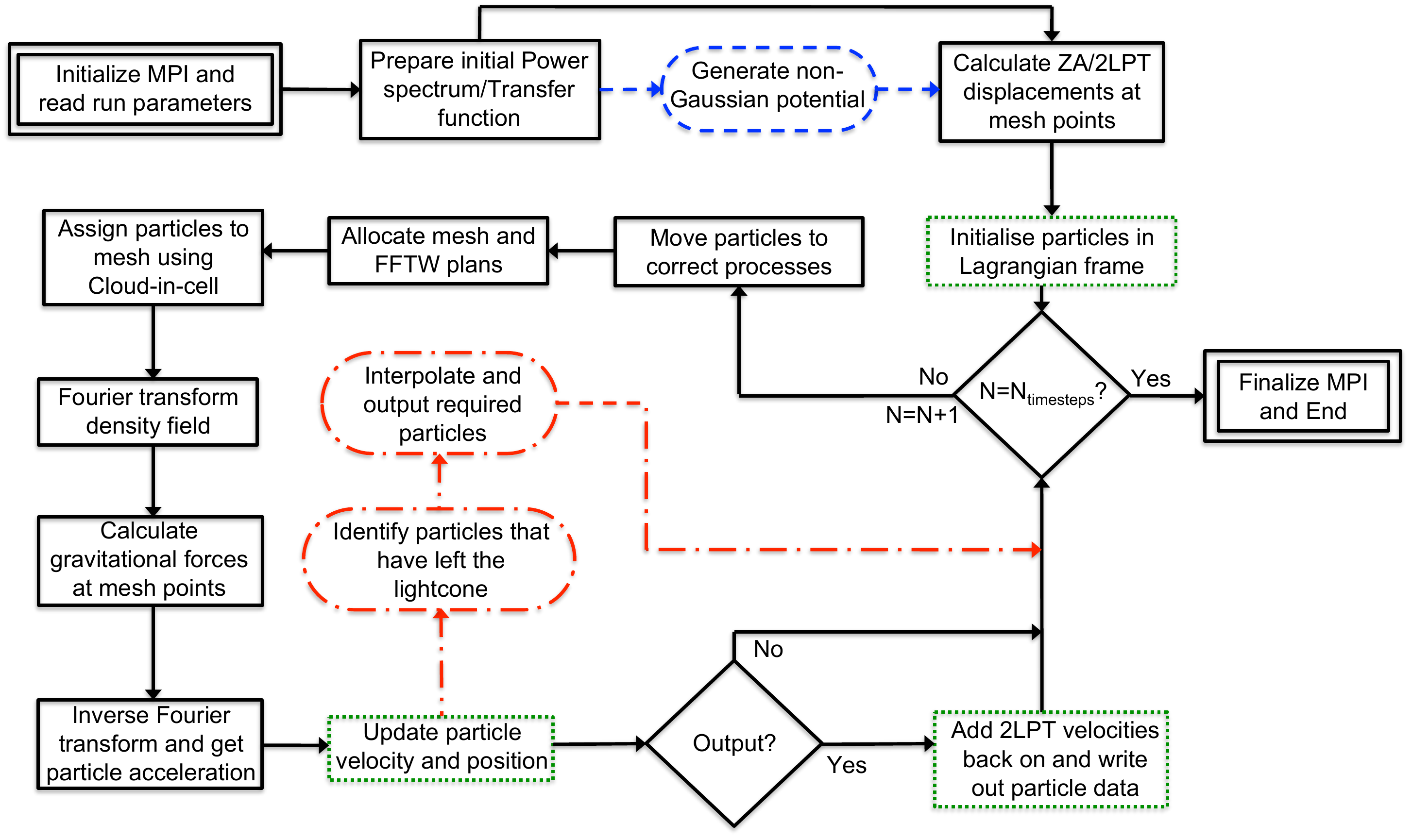}
  \caption{A flowchart detailing the steps {\sc l-picola} takes in generating a dark matter realisation from scratch. The green, dotted boxes indicate where the COLA algorithm is applied, differentiating {\sc l-picola} from a standard PM code. The blue, dashed box indicates where the inclusion of primordial non-Gaussianity changes the code structure. The red, dot-dash boxes highlight areas where the code differs depending on whether we are running snapshot or lightcone simulations.}
  \label{PICOLAchart}
\end{figure*}

Fig. \ref{PICOLAchart} shows a simple step-by-step overview of how {\sc l-picola} works. The different coloured boxes highlight areas where the structure of the code actually changes depending on how it is compiled. The blue box shows where the different types of non-Gaussianity can be included. The red boxes show where significant algorithmic changes occur in the code if lightcone simulations are requested. These will we detailed in the following sections, along with an explanation of how we parallelise the COLA method.

{\sc l-picola} is publicly available under the GNU General Public License at \url{https://cullanhowlett.github.io/l-picola}.

\section{Parallelisation}
In this section we will detail the steps we have taken to parallelise the COLA method. All parallelisation in the code uses the Message Passing Interface (MPI) library\footnote{This software package can be downloaded at \url{http://www.open-mpi.org/}}. See \cite{Pacheco1997} for a comprehensive guide to the usage and syntax of MPI. In the following subsections we provide an overview to the parallelisation and detail the three main parallel algorithms in the code: parallel Cloud-in-Cell interpolation, parallel FFT's and moving particles between processors.

\subsection{Parallelisation Overview}
Parallelisation of {\sc l-picola} has been performed with the goal that each processor can run a small section of the simulation whilst needing minimal knowledge of the state of the simulation as a whole. We have separated both the mesh and particles across processors in one direction. In this way each processor gets a planar portion of the mesh, and the particles associated with that portion. We have tried to balance the load on each processor as much as possible whilst adhering to the fact that each processor must have an integer number of mesh cells in the direction over which we have split the full mesh. 

This process is enabled by use of the publicly available FFTW-MPI libraries, which also serve to perform the Fast Fourier Transforms when the mesh is split over different processors\footnote{These are included in the FFTW package mentioned previously}. In a simulation utilising $N_{p}$ processors and consisting of a cubic mesh of size $N_{mesh}^{3}$, each processor gets $(\lceil N_{m}/N_{p} \rceil)$ slices of the mesh where each slice consists of $N_{m}\times 2(N_{m}/2+1)$ cells. The extra $2N_{m}$ cells in each slice are required as buffer memory for the FFTW routines. Depending on the ratio of $N_{m}$ to $N_{p}$ this may give too many slices in total, so then we work backwards, removing slices until the total number of slices is equal to $N_{m}$. 

The number of particles each processor has is related to the number of mesh cells on that processor as each processor only requires knowledge of any particles that interact with its portion of the mesh. Hence, as the particles are originally spaced equally across the mesh cells, each processor initially holds $N^{3}/N_{m}$ particles, multiplied by the number of slices it has.

\subsection{Parallel Cloud-in-Cell}

\begin{figure}
\centering
\includegraphics[width=84mm]{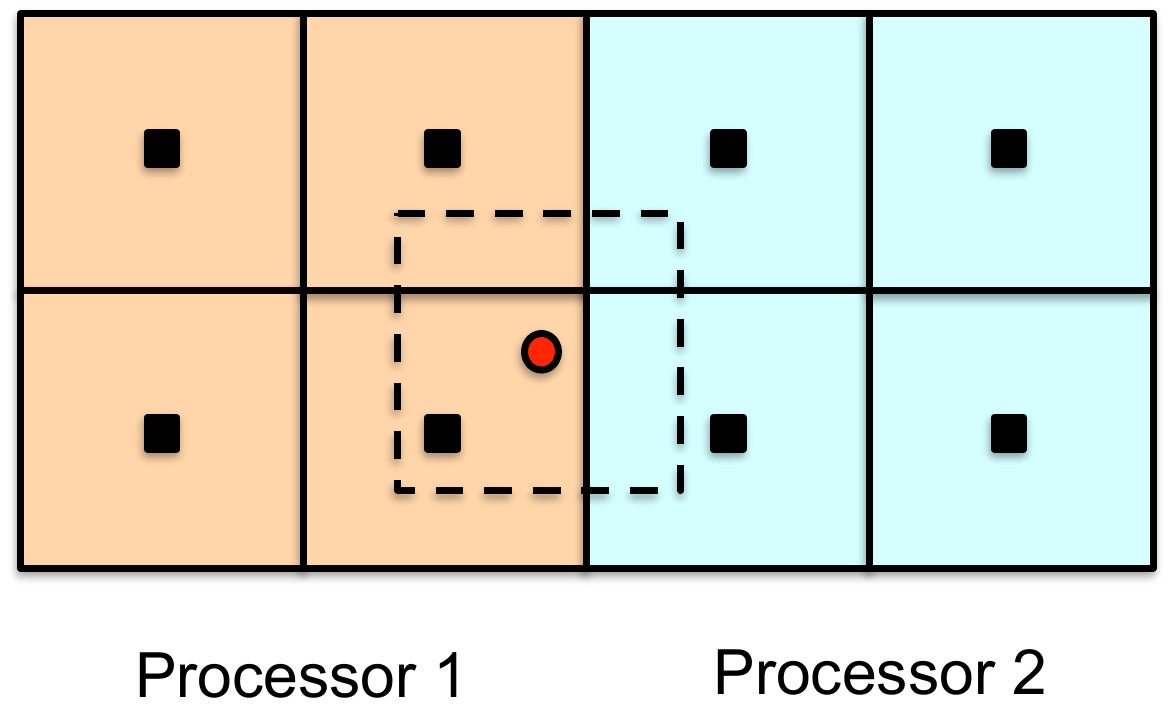}
  \caption{A visual representation of the 2-D Cloud-in-Cell algorithm. The particle (red point) is shared across the four nearest mesh cells with weight given by the percentage of the particle's `cloud' (dashed line) that overlaps the mesh cell. However, in {\sc l-picola} these mesh cells may not be on the same processor as the particle. This is corrected by assigning extra slices of the mesh to the `right-hand' edge of processor $i$ which are then transferred and added to the left most slice on processor $i+1$.}
  \label{cloudincell}
\end{figure}

As each processor only contains particles which belong to the mesh cells it has, and our interpolation assigns particles to the mesh by rounding down to the nearest mesh cell, the density assignment step proceeds as per the standard Cloud-in-Cell interpolation method, except near the `left-hand' edge of the processor. Here the density depends on particles on the preceding processor. Figure \ref{cloudincell} shows a 2-D graphical representation of this problem.

In order to compensate for this we assign an extra mesh slice to the `right-hand' edge of each processor. This slice represents the leading slice on the neighbouring processor and by assigning the particles to these where appropriate and then transferring and adding the `slices' to the appropriate processors, each portion of the mesh now contains an estimate of the density which matches the estimate as if all the mesh were contained on a single processor. 

It should also be noted that a reverse of this process must also be done after calculating the forces at each mesh point, as the displacement of a particle near the edge of a processor is reliant on the force at the edge of the neighbouring processor.

\subsection{Parallel FFT's}

To take the Fourier transform of our mesh once it is split over many processors we use the parallel FFTW-MPI routines available alongside the aforementioned FFTW libraries. This is intimately linked to the way in which the particles and mesh are actually split over processors and routines are provided in this distribution that enable us to perform this split in the first place. 

The FFTW routines use a series of collective MPI communications to transpose the mesh and perform a multi-dimensional real-to-complex discrete Fourier transformation of the density, assigning the correct part of the transformed mesh to each processor. In terms of implementing this, all that is required is for us to partition the particles and mesh in a way that is compatible with the FFTW routines, create a FFTW plan for the arrays we wish to transform and perform the Fourier transformation once we have calculated the required quantity at each mesh point. The FFTW libraries perform all MPI communications and operations internally.

\subsection{Moving Particles}

One final modification to the Particle-Mesh algorithm is to compensate for the fact that, over the course of the simulation, particles may move outside the physical area contained on each processor. Their position may now correspond to a portion of the mesh that the processor in question does not have. As such, after each timestep we check to see which particles have moved outside the processor boundaries and move them to the correct processor. This is made particularly important as the COLA method converges in very few timesteps, meaning the particles can move large distances in the space of a single timestep. 

In the case where we have a high particle density or small physical volume assigned to each processor, a single particle can jump across several neighbouring processors in a single timestep. So, when moving the particles, we iterate over the maximum number processors any single particle has jumped across. However the number of particles that need to be moved is unknown \textit{a priori} and so to be conservative and make sure that we do not overload the buffer memory set aside for the transfer, not all the particles that are moving are transferred simultaneously (i.e. via a collective MPI-Alltoall command). Rather, all the particles that have moved from processor N to N$\pm$1 are moved first then all the particles that have moved from processor N to N$\pm$2 are transferred. Although this requires iterating over the particles on processor N multiple times, in the majority of cases there are no particles moving to any processors beyond N$\pm$1 and so only one iteration is required.

\begin{figure}
\centering
\includegraphics[width=84mm]{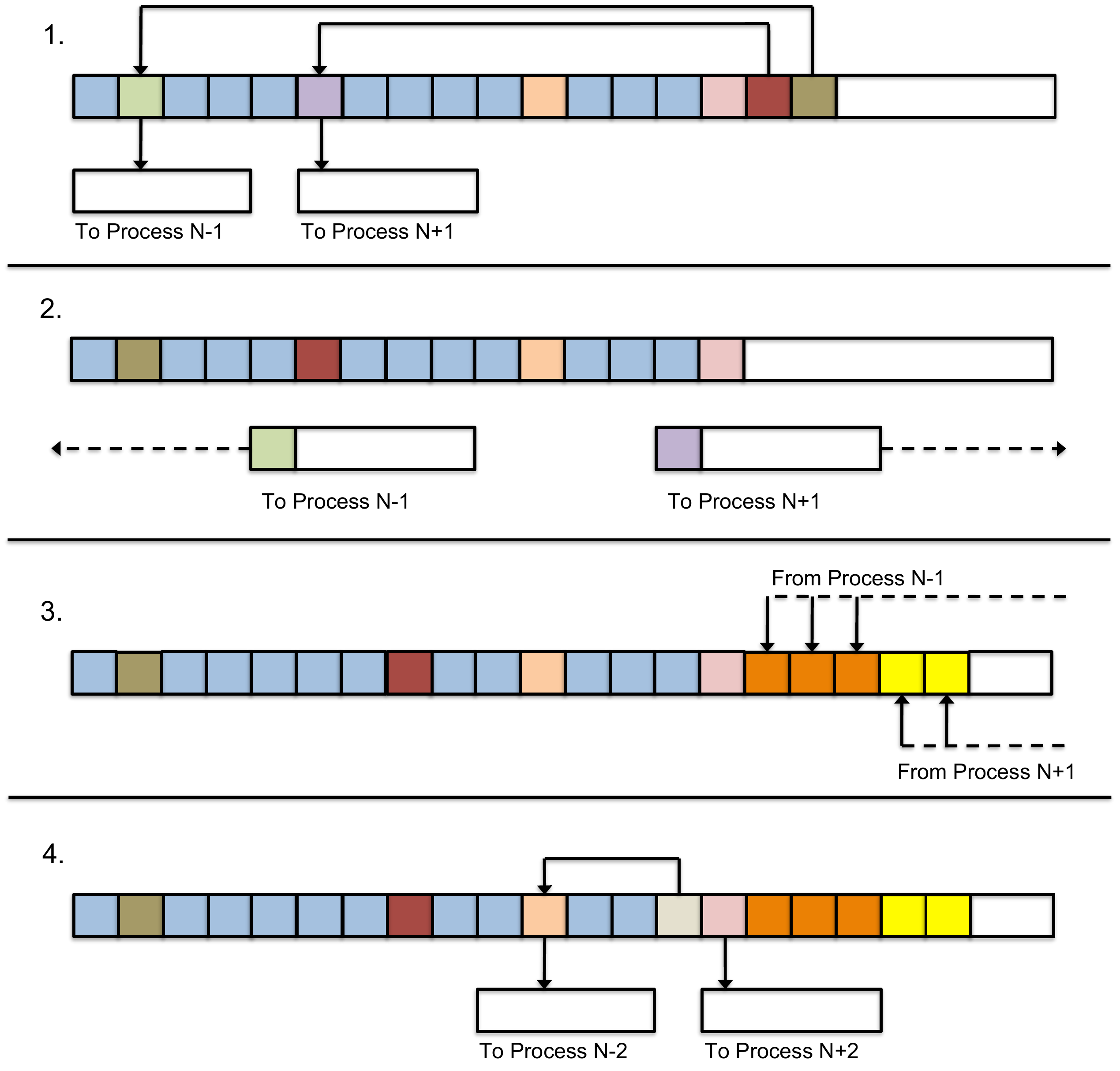}
  \caption{A four stage `memory schematic' of how we move particles between processors in between timesteps, conserving as much memory as possible. First we identify those particles which need moving to the neighbouring processors and move them to a temporary buffer. We then move particles from the end of the particle structure to overwrite the particle we no longer need to keep. Finally we perform a send and receive operation, sending the particles in the buffer to the neighbouring processors and receiving particles from those processors into the end of the particle structure. This algorithm is repeated up to the maximum number of processors a particle has moved across.}
  \label{moveparticles}
\end{figure}

As the simulation progresses the particles will not remain homogeneously spread over the processes, so we assign additional buffer memory to each processor to hold any extra particles it acquires. This is utilised during the moving of the particles and all particles a processor receives are stored in this buffer. However, in order to make sure this buffer is not filled too quickly we also use the fact that each processor is likely to lose some particles. When a particle is identified as having left a particular processor the particle is moved into temporary memory and the gap is filled with a particle from the end of the processors main particle memory. In this way we collect all remaining particles together before moving the new particles across, ensuring a contiguous, compact particle structure. This is shown in Figure \ref{moveparticles}.

\section{Generating Initial Conditions}
In order to allow {\sc l-picola} to run a simulation from scratch we have integrated an initial conditions generator into the code. This means that we can simply store the first and second order Lagrangian displacements for each particle as they are calculated rather than assume some initial positions for the particles and reconstruct them. We use the latest version of the parallelised 2LPTic code\footnote{A parallelised version of the code including primordial non-Gaussianity can be found at \url{http://cosmo.nyu.edu/roman/2LPT/.}} \citep{Scoccimarro1998, Scoccimarro2012} to generate the initial conditions, with some modifications to allow a more seamless combination of the two codes, especially in terms of  parallelisation. For compatibility with {\sc l-picola} we have removed the warm dark matter and non-flat cosmology options from the 2LPT initial conditions generator, though these are improvements that could easily be added in the future. The particles are initially placed uniformly, in a grid pattern throughout the simulation volume, so rather than creating the particles at this stage, we also conserve memory by only generating the 2LPT displacements at these points and creating the particles themselves just before timestepping begins.

Because of this addition, {\sc l-picola} can be used very effectively to create the initial conditions for other N-Body simulations, as well as evolving the dark matter field itself. In fact in a single run we can output both the initial conditions and the evolved field at any number of redshifts between the redshift of the initial conditions and the final redshift, which allows easy comparison between PICOLA and other N-Body codes.

A final point is that because the 2LPT section is based on the latest version of the 2LPTic code, we are also able to generate, and then evolve, initial conditions with local, equilateral, orthogonal or generic primordial non-Gaussianity. Local, equilateral and orthogonal non-gaussianity can be added simply by specifying the appropriate option before compilation and providing a value for $f_{NL}$. We can also create primordial non-Gaussianity for any generic bispectrum configuration using a user-defined input kernel, following the formalism in the Appendix of \cite{Scoccimarro2012}.

\section{Lightcone}
The final large modification we have made to the code, and one which will be very useful for future large scale structure surveys, is the ability to generate lightcone simulations in a single run, as opposed to running a large number of snapshots and piecing them together afterwards.

Snapshot simulations, generated at some effective redshift, have been widely used in the past to calculate the covariance matrix and perform systematic tests on data \citep{Manera2013, Manera2015}. However, as future surveys begin to cover larger and larger cosmological volumes with high completeness across all redshift ranges it is no longer good enough to produce a suite of simulations at one redshift. Lightcone simulations mimic the observed clustering as a function of redshift and so introduce a redshift dependence into the covariance matrix. On top of this, once a full redshift range has been simulated we can apply identical cuts to the mock galaxy catalogues and the data. As such, in the case when we make measurements at multiple effective redshifts with a single sample, we may need less simulations in total, especially if multiple runs would be required to produce the snapshots at multiple redshifts.

\begin{figure}
\centering
\includegraphics[width=84mm]{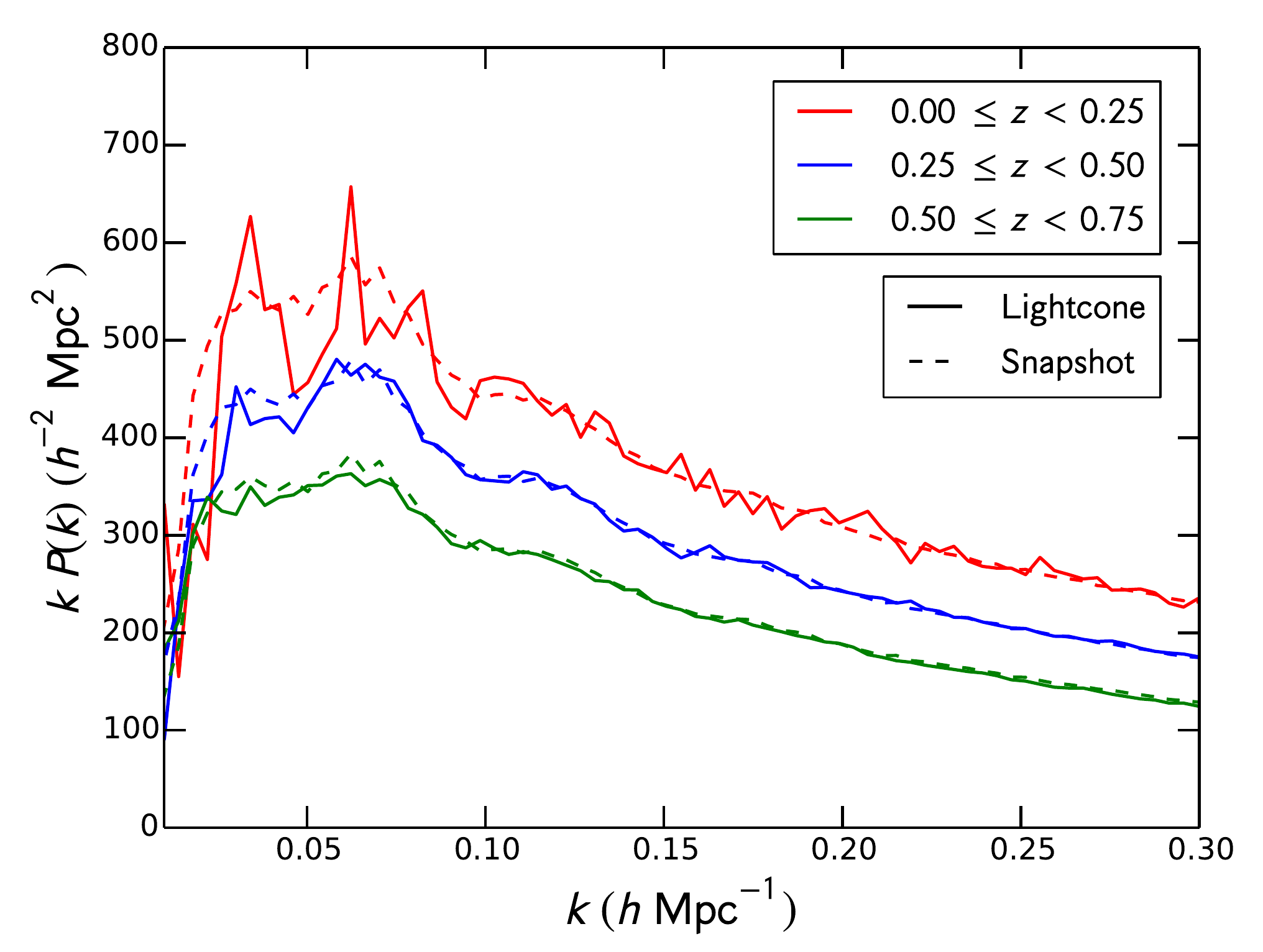}
  \caption{The power spectra, measured using the estimator of \cite{Feldman1994}, of different redshifts slices within the same {\sc l-picola} lightcone simulation (solid). This is compared to a snapshot simulation at the effective redshift of the slice. We see that, as expected, the clustering is much stronger (and the power spectrum amplitude much higher) at lower redshifts and there is good agreement on linear scales between the snapshot and lightcone power spectra.}
  \label{lightcone_pk}
\end{figure}

Figure \ref{lightcone_pk} demonstrates the effect of simulating a lightcone using the power spectrum. We populate a $(2\gpcoh)^{3}$ box with $512^{3}$ particles, place the observer at (0,0,0), and using a flat, $\Omega_{m,0}=0.25$ cosmology (all other parameters match our fiducial cosmology), simulate an eighth of the full-sky out to a maximum redshift of 0.75. The power spectrum is then calculated using the method of \cite{Feldman1994} for three redshift slices between 0.0, 0.25, 0.5 and 0.75, using a random, unclustered catalogue to capture the window function. As expected we see a significant evolution of the clustering as a function of redshift that would not be captured in a single snapshot simulation. The overall clustering amplitude increases as we go to lower redshifts with additional non-linear evolution on small scales at later times. 

To further compare the clustering of this lightcone simulation with the expected clustering, we overlay the power spectrum from a snapshot simulation at the effective redshift of each lightcone slice. We define the effective redshift of each slice, bounded by the redshifts $z_{1}$ and $z_{2}$, using the formulation of \cite{Tegmark1997} where 
\begin{equation}
z_{eff} = \frac{\int(n(z)P_{FKP}w_{FKP})^{2}zdV}{\int(n(z)P_{FKP}w_{FKP})^{2}dV}.
\end{equation}
For our simulations the number density, $n(z)$, is constant and the weighting factors, $P_{FKP}$ and $w_{FKP}$ \citep{Feldman1994} cancel. This in turn reduces the effective redshift to
\begin{equation}
z_{eff} = \frac{\int_{z_{1}}^{z_{2}} \frac{r^{2}(z)}{H(z)}zdz}{\int_{z_{1}}^{z_{2}}\frac{r^{2}(z)}{H(z)}dz}
\end{equation}
where 
\begin{equation}
r(z) = c\int_{0}^{z}\frac{dz'}{H(z')}
\end{equation}
is the comoving distance, $c$ is the speed of light and $H(z)$ is the Hubble parameter.

We can see good agreement on all scales between the snapshot and lightcone power spectra for each of the redshift slices. The window function causes noise on the largest scales, especially for the lowest volume slice, however the redshift-dependent amplitude is captured very well within a single lightcone simulation.

In the following subsections we will provide a detailed description of how lightcone simulations are produced in {\sc l-picola}, test the accuracy of our implementation and also looking at how we can replicate the simulations volume to fill the full lightcone during run-time.

\subsection{Building Lightcone Simulations} \label{sec:build_lightcone}

In order to simulate the past lightcone, we require the properties of each particle in the simulation at the moment when it leaves a lightcone shrinking towards the observer. As has been done in several studies \citep{Fosalba2013, Merson2013}, we can interpolate these particle properties using a set of snapshot simulations, however this requires significant post-processing and more storage space than generating a lightcone simulation at run-time. As such, in order to provide a useful tool for future cosmology surveys, we have implemented the latter into {\sc l-picola}.

\begin{figure}
\centering
\includegraphics[width=84mm]{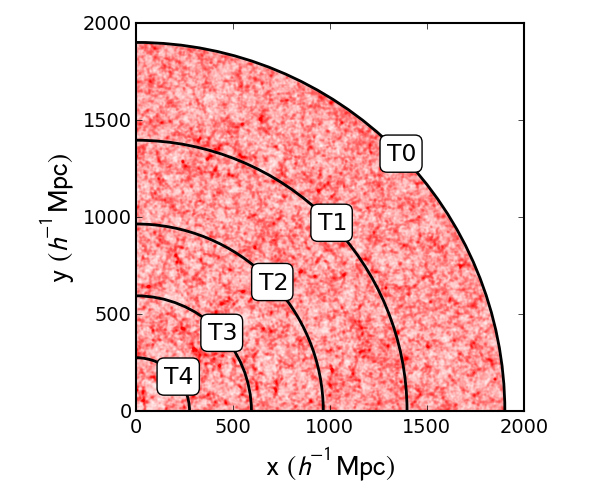}
  \caption{A $50\mpcoh$ slice of a {\sc l-picola} dark matter field simulated on the past lightcone with an observer situated at the (0,0,0). As the lightcone shrinks with each timestep (the lightcone radius is denoted by the black lines) we only output the particles that have left the lightcone that timestep, with their position interpolated to the exact point at which they left. This means that the particles shown in the diagram were output in stages with the particles between lines T0 and T1 output first. Between output stages the particles evolve as normal, resulting in clustering that is dependent on the distance from the observer.}
  \label{lightcone}
\end{figure}

This is done as follows: The user specifies an initial redshift, at which we begin the simulation, and an origin, the point at which the observer sits. Each of the output redshifts is then used to set up the timesteps we will use in the simulation, with the first output denoting the point at which we start the lightcone and the final output corresponding to the final redshift of the simulation. Any additional redshifts in between these two can be used to set up variable timestep sizes. If we imagine the lightcone as shrinking towards the origin as the simulation progresses, then for every timestep between these two redshifts we output only those particles that have left the lightcone. This is shown pictorally in Figure \ref{lightcone}. 

Mathematically, it is simple to identify whether the particle should be output between timesteps $i$ and $i+1$ by looking for particles which satisfy both
\begin{equation}
R_{p,i} \le R_{L,i}
\end{equation}
and
\begin{equation}
R_{p,i+1} > R_{L,i+1},
\end{equation}
where $R_{p,i}$ is the comoving distance between the particle and the lightcone origin at scale factor $a_{i}=1/(1+z_{i})$ and $R_{L,i}$ is the comoving radius of the lightcone at this time. 

\begin{figure*}
\centering
\subfloat{\includegraphics[width=0.5\textwidth]{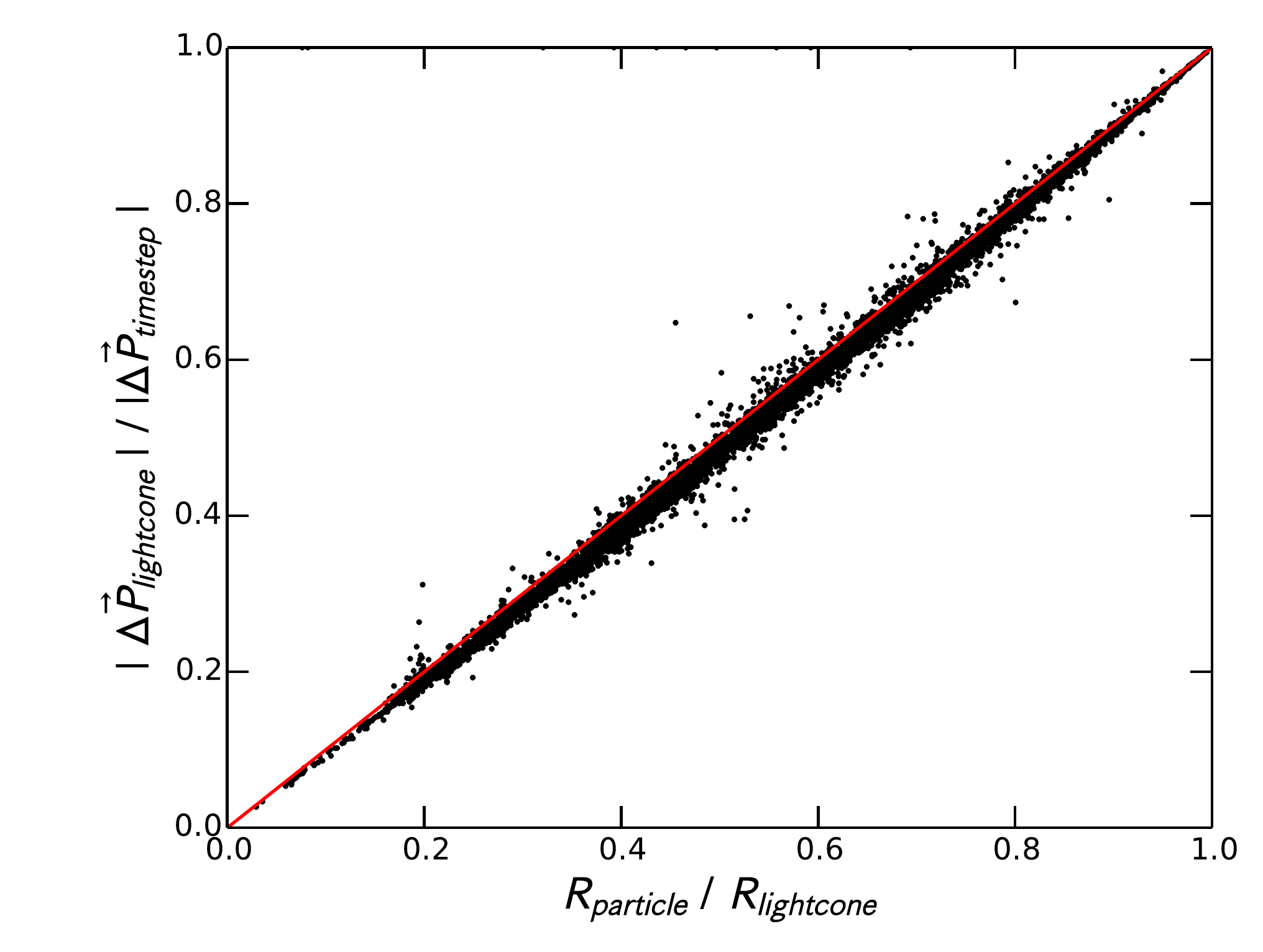}}
\subfloat{\includegraphics[width=0.5\textwidth]{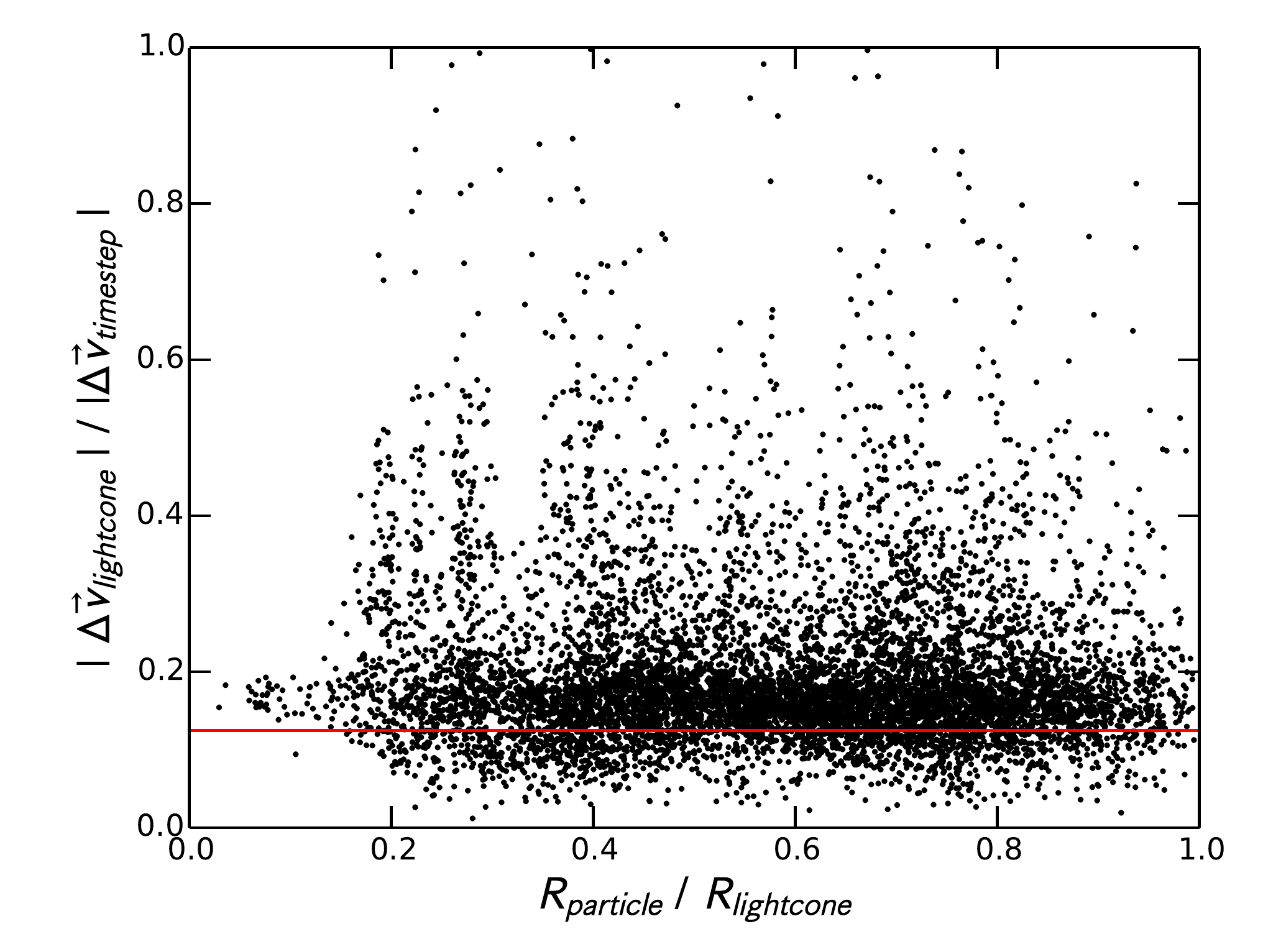}}
\caption{The difference between the lightcone and snapshot positions (left) and velocities (right) of particles output between $z=0.0$ and $z=0.09375$ as a function of the distance to the observer, which is equivalent to the output time. We compute the magnitude of the difference vector between the lightcone and $z=0.0$ snapshot statistics, normalised by the difference between the $z=0.0$ and $z=0.09375$ snapshots. The solid red line shoes the expected trend based on the fact that we output particles at the time they exit the lightcone, but do not interpolate the velocity. For the latter we expect, for each direction, $v_{a+1/2} \approx (v_{a+1}+v_{a})/2$.}
\label{picola_interp_vel}
\end{figure*}

However, we really wish to output a given particle at the exact moment it satisfies the equation,
\begin{equation}
R_{p}(a_{L}) = R_{L}(a_{L}),
\label{eq:lightcone}
\end{equation}
From the {\sc cola} method we have 
\begin{equation}
R_{p}^{2}(a_{L})= |\boldsymbol{r}_{i} - \boldsymbol{r}_{0}+ \boldsymbol{v}_{i+1/2}\Delta a_{2}+ \Delta D_{1}\boldsymbol{\Psi}_{1} + \Delta D_{2}\boldsymbol{\Psi}_{2} |^{2}
\end{equation}
where $\boldsymbol{r}_{0}$ is the position of the lightcone origin and the `$\Delta$' terms are dependent on the value of $a_{L}$. The comoving lightcone radius at $a_{L}$ is simply the comoving distance 
\begin{equation}
R_{L}(a_{L})=c\int_{a_{L}}^{1}\frac{da}{a^{2}H(a)}.
\end{equation}
Equating these should allow us to solve for $a_{L}$. Once this is known we can calculate the properties of each particle we wish to output. However, this equation cannot be solved analytically and so requires us to numerically solve it for each individual particle that we wish to output. This would be prohibitively time-consuming and instead we approximate the solution by linearly interpolating both the lightcone radius and the particle position between the times $a_{i}$ and $a_{i+1}$. Substituting the linear interpolation into Eq.(\ref{eq:lightcone}) and rearranging we find 
\begin{equation}
a_{L} \approx a_{i}+\frac{(a_{i+1}-a_{i})(R_{L,i}-R_{p,i})}{(R_{p,i+1}-R_{p,i})-(R_{L,i+1}-R_{L,i})}.
\label{eq:lightconeinterp}
\end{equation}
This is trivial to calculate as we already need to know $R_{p,i}$ and $R_{p,i+1}$ in order to update the particle during timestepping anyway, and $R_{L,i}$ and $R_{L,i+1}$ are needed to identify which particles have left the lightcone in the first place. In fact the whole procedure can be performed with minimal extra runtime, as we simply modify the `Drift' part of the code. The only extra computations are to check the particle's new position against the lightcone and interpolate if necessary. Once we know the exact time the particle left the lightcone we can update the particle's position, using Eq.(16), to the position it had when it left the lightcone and output the particle. 

In {\sc l-picola} lightcone simulations we do not interpolate the velocity, using instead the velocity at time $a_{i+1/2}$. We make this choice as it mimics the inherent assumption of the Kick-Drift-Kick method, that the velocity is constant between $a_{i}$ and $a_{i+1}$. To properly interpolate the velocity in the same way as the particle position would require us to evaluate the velocity at times $a_{i}$ and $a_{i+1}$ which in turn would require us to measure the particle density at half timestep intervals. One could also imagine assuming that the non-linear velocity is constant and interpolating the ZA and 2LPT velocities (which must be added back on before outputting to move back to the correct reference frame). However, we find that the assumption of constant velocity between $a_{i}$ and $a_{i+1}$ is a reasonable one.

To test the numerical interpolation against the analytic expectations, and provide a graphical representation of the particle positions and velocities output during lightcone simulations, we compare particles output during the final timestep of a lightcone simulation to the same particles output from snapshot simulations evaluated at the beginning and end of that timestep (the corresponding redshifts are $z=0.0$ and $z=0.09375$ in this case). The particles are matched based on a unique identification number which is assigned when the particle are created and as such is consistent between the three simulations.

For both the particle positions and velocities we look at the difference between the lightcone properties and the properties of the $z=0.0$ snapshot, normalised by the same difference between the $z=0.09375$ and $z=0.0$ snapshots. We plot this in Figure \ref{picola_interp_vel} as a function of the distance from the observer (also normalised, using the comoving distance to $z=0.09375$). If we were to interpolate the particle positions after runtime using the two snapshots we would expect the particles to lie exactly on the diagonal in Figure \ref{picola_interp_vel}. We find that the particle positions interpolated \textit{during} the simulation also lie close to the diagonal, which validates the accuracy of our numerical interpolation. The small scatter in both of these plots is due to floating-point errors and the normalisation in the particle positions. Particles that do not move much between the two snapshots will have a normalisation close to zero, which in turn makes our choice of plotting statistic non-optimal. The particle velocities show no trend as a function of distance to the observer or when they were output. In this case the velocities in each direction are all situated close to the mid point between the two simulations . This validates the Kick-Drift-Kick assumption, that the velocities evolve approximately linearly between two timesteps, such that the velocity at time $a_{i+1/2}$ is half way between that at time $a_{i}$ and $a_{i+1}$, although there is some scatter and offset due to the true non-linear nature of the velocity.

\subsubsection{Interpolation Accuracy}

\begin{figure}
\centering
\includegraphics[width=84mm]{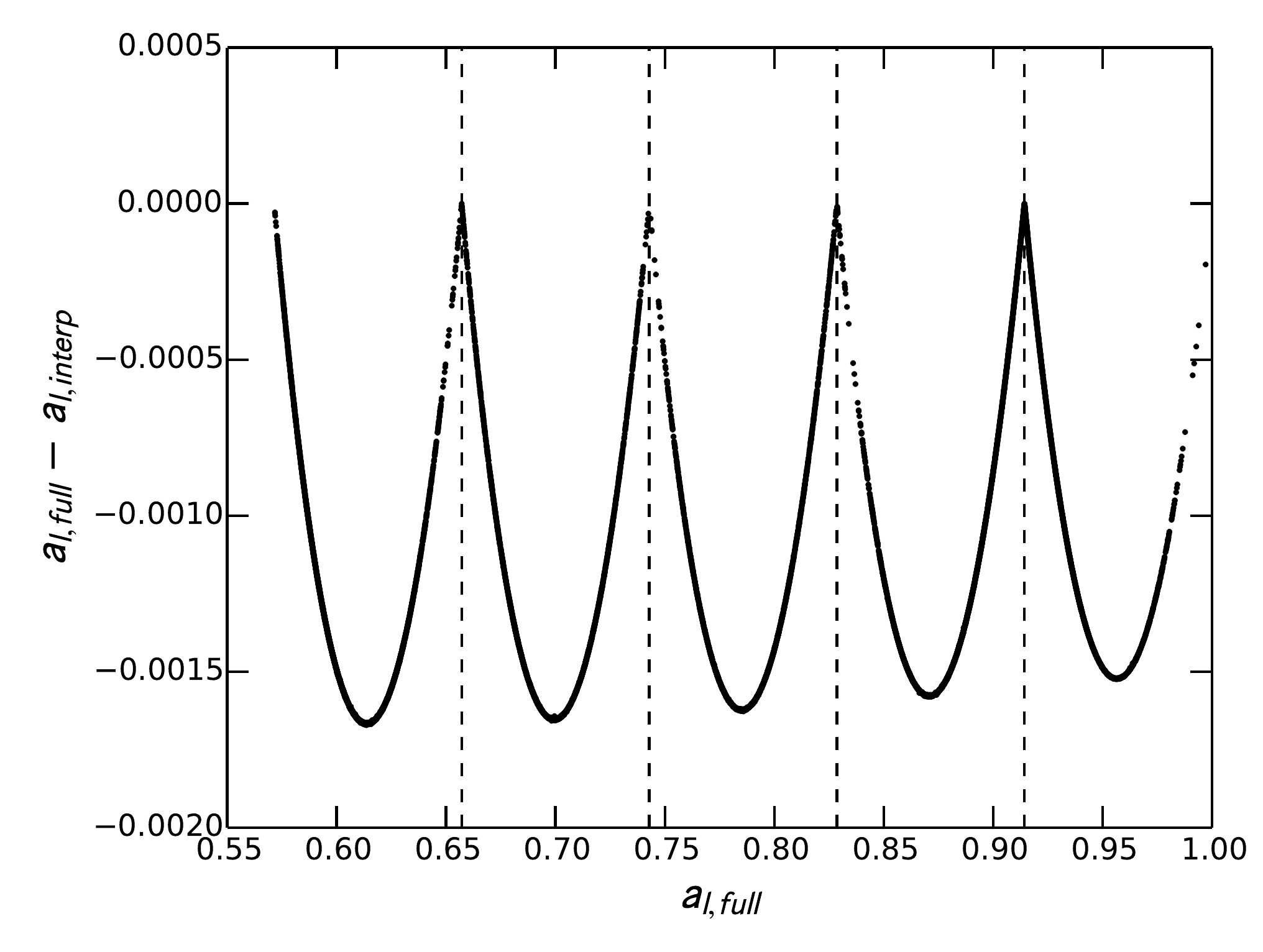}
  \caption{A plot showing the accuracy of using linear interpolation to get the time a particle leaves the lightcone. For a subset of particles, we plot the difference between the full numerical solution of $a_{L}$ and the value recovered using Eq.(\ref{eq:lightconeinterp}), as a function of the true scale factor. The dashed lines show the scale factor at which we evaluate the timesteps of the simulations, and hence know the exact positions of the particles and the lightcone radius.}
  \label{picola_interp_a}
\end{figure}

On top of comparing the numerical interpolation during runtime to the analytic interpolation between two snapshots, we also check the assumption that we can use linear interpolation between two timesteps \textit{at all}.  As mentioned previously, the exact time the particle leaves the lightcone, which we'll call $a_{L,full}$, is given by numerically solving Eq.(\ref{eq:lightcone}), but solving this for each particle is extremely time consuming and so we linearly interpolate instead.  To test this we find the exact solution for a subset of the particles in the $L=2\gpcoh, N=512^{3}$ simulation and compare this to the approximate solution, $a_{L,interp}$. This is shown in Figure \ref{picola_interp_a}. We find that the linear interpolation slightly overestimates the value of $a_{L}$, with a common trend across all timesteps, however this effect is less than 0.5\% across all times for this simulation. The two solutions agree almost perfectly close to the timestep boundaries, denoted by the dashed vertical lines. This is because the particle positions and lightcone radius are known exactly at these points. Further away from the timestep boundaries inaccuracies are introduced as the assumption that the particle position and lightcone radius are linear functions of the scale factor is less accurate.

\begin{figure}
\centering
\includegraphics[width=84mm]{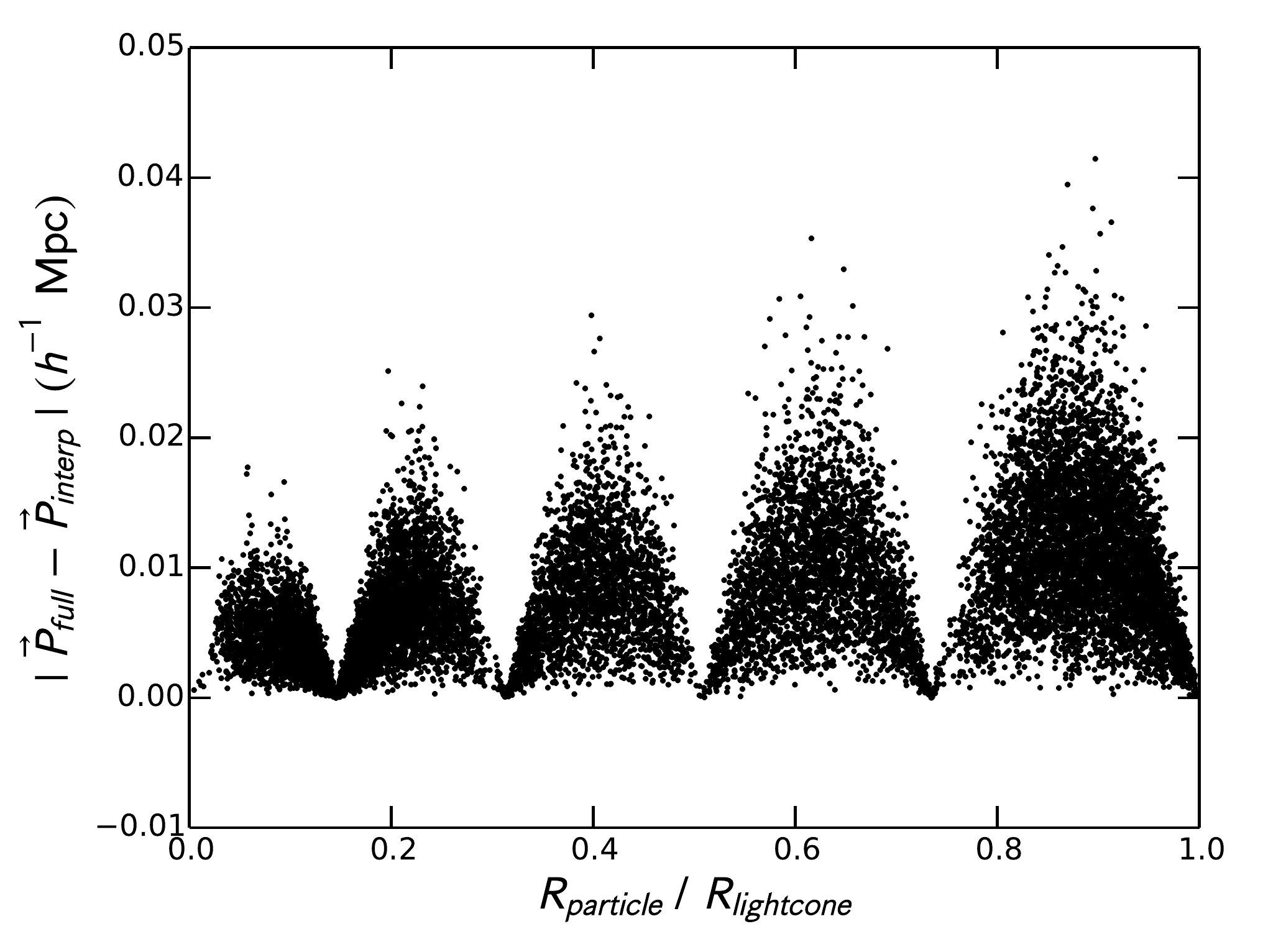}
  \caption{A plot of the difference between the positions of a subset of particles when using the full numerical solution of $a_{L}$ and those recovered using Eq.(\ref{eq:lightconeinterp}), as a function of the distance from the observer, normalised by the maximum lightcone radius of the simulation. We plot the magnitude of the difference vector between the two methods. We see good agreement, with a maximum difference of $\sim 60 h^{-1}\,{\rm kpc}$, across all scales.}
  \label{picola_interp_pos2}
\end{figure}

We can further quantify the reliability of the linear interpolation by looking at the positions of the particles output in both these simulations. This is shown in Figure \ref{picola_interp_pos2}, where we plot the difference in particle position (we take the magnitude of the difference vector) as a function of the distance between the particle and the observer, normalised to the maximum lightcone radius for the simulation. We can see that the linear interpolation is indeed very accurate, and even at large radii, where the comoving distance between timesteps is largest, the particle positions are equivalent to within $0.06\mpcoh$. This is well below the mesh scale of this simulation, and is subdominant compared to the errors caused by the finite mesh size and the large timesteps.

\subsection{Replicates}

\begin{figure}
\centering
\includegraphics[width=84mm]{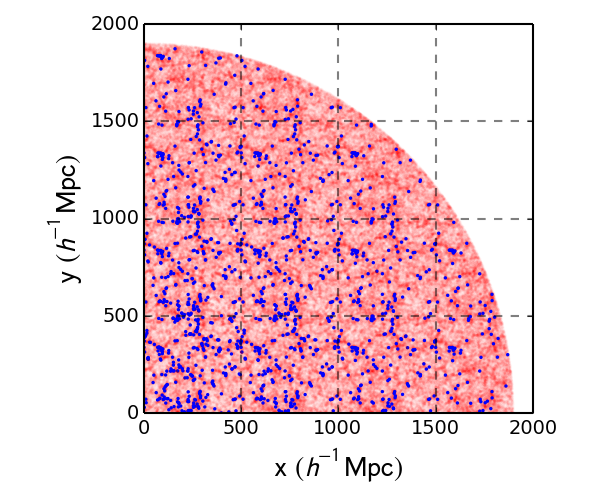}
  \caption{A {\sc l-picola} lightcone simulation showing obvious replicates. In this rather extreme case we run a similar simulation to that shown in Figure \ref{lightcone} but using 64 times less particles and in a volume 64 times smaller. We then replicate the box 64 times at runtime as shown by the dashed lines (we only show 1 replicate in the z direction). To aid visualisation we also over plot the halos recovered from this simulation using a Friends-of-Friends algorithm.}
  \label{lightcone_replicate}
\end{figure}

On top of the lightcone interpolation we have accounted for the fact that lightcones built from snapshot simulations often replicate the simulation output to reach the desired redshift. {\sc l-picola} has the ability to replicate the box as many times as required in each direction during runtime. This is done by simply modifying the position of each particle as if it was in a simulation box centred at some other location. In this way we can build up a large cosmological volume whilst still retaining a reasonable mass and force resolution. However it is important to note that this can have undesired effects on the power spectrum and covariance matrix calculated from the \textit{full} replicated simulations volume, which will be detailed subsequently. Figure \ref{lightcone_replicate} highlights the replication process. Here we run a similar lightcone to that used in Figure \ref{lightcone}, however the actual simulation contains 64 times less particles in a volume 64 times smaller and is replicated 64 times. In this way we can cover the full volume and mass range required but the CPU and memory requirements are much smaller. To help identify the replication we have used the Friends-of-Friends algorithm \citep{Davis1985} to group the particles into halos and plotted the centre-of-mass position of each halo. This results in obvious points where the same halo is reproduced after more particles have accreted onto that halo, and the halo has evolved in time. 

\subsubsection{Effects of Replication on the Power Spectrum}

The downside of the replication procedure is that in repeating the same structures we are not be sampling as many independent modes as would be expected from an unreplicated simulation of the same volume. Rather we are just sampling the same modes multiple times. This affects both the power spectrum and the covariance matrix. To test the effects of replication we use a set of 500 lightcone simulations, containing $512^{3}$ particles in a box of edge length $1024\mpcoh$. We then compare this to another set of 500 simulations with $256^{3}$ particle in a $(512\mpcoh)^{3}$ box, which is then replicated 8 times. We calculate the power spectra for both using the method of \cite{Feldman1994}, in bins of $\Delta k=0.008 h\,{\rm Mpc}^{-1}$, estimating the expected overdensity from the total number of simulation particles and the box volume. This works for the lightcone simulations as the maximum lightcone radius is larger than the diagonal length of the cubic box, such that the simulation still fills volume.

\begin{figure}
\centering
\includegraphics[width=84mm]{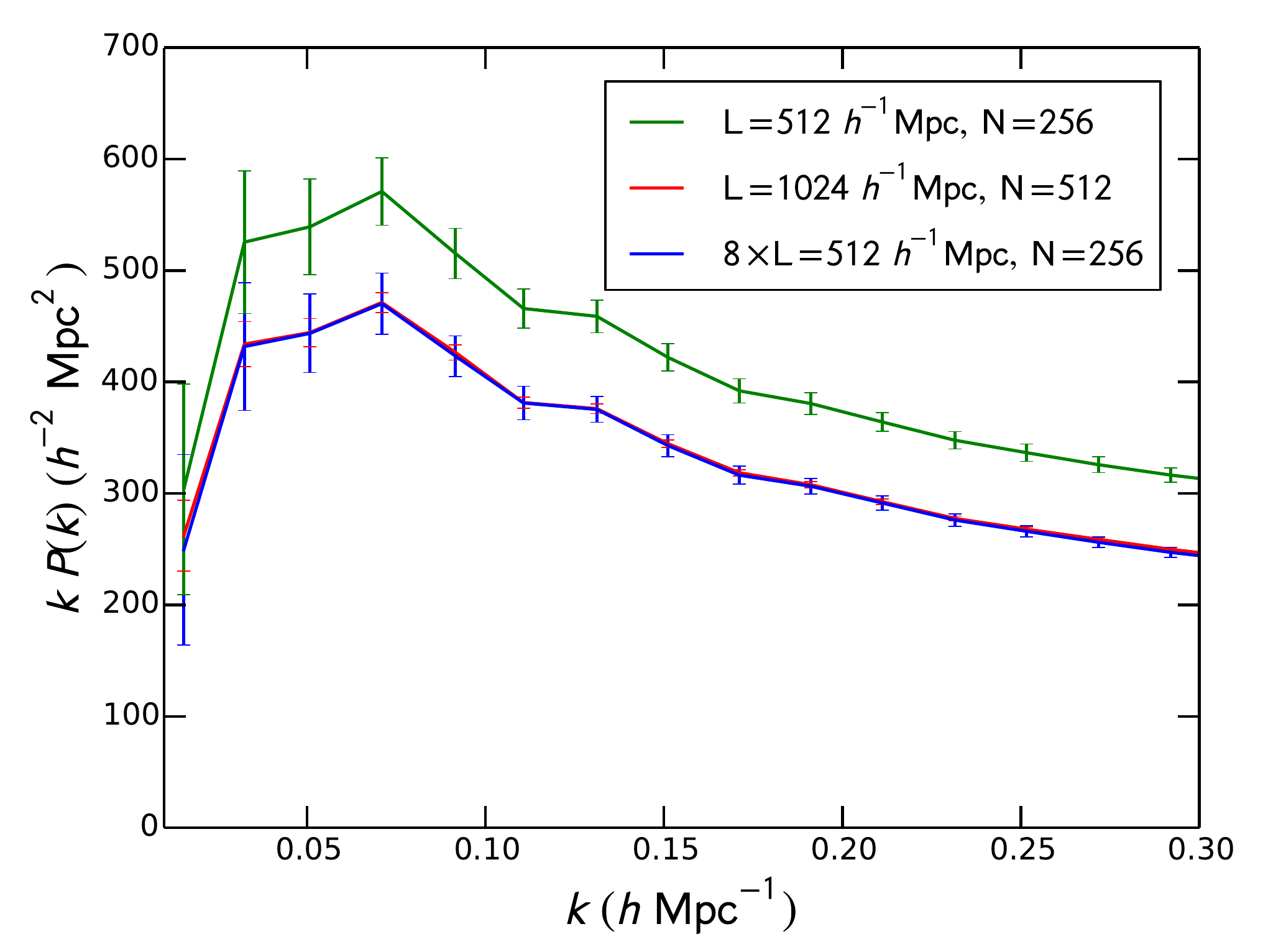}
\caption{Plots showing the effect of replication on the estimated power spectrum using sets of $512^{3}$ particle lightcone simulations in a $1024\mpcoh$ box and $256^{3}$ particle lightcone simulations in a $512\mpcoh$ box. The lines correspond to the average power spectra from 500 independent realisations and the errors are those on a single realisation calculated from the diagonal of the covariance matrix constructed from the 500 realisations. The blue line represents the average power spectrum when we replicate the $256^{3}$ particle simulation 8 times so that it has the same volume and number of particles as the larger simulation, and as expected is virtually indistinguishable from the large, unreplicated simulation. The amplitude of power spectra match the order of the legend.}
\label{replicate_pk}
\end{figure}

The average power spectra are shown in Figure \ref{replicate_pk}, where the errors come from the diagonal elements of the covariance matrix and are those for a single realisation. As the simulations are periodic in nature we expect the power spectra for the two box sizes to be  almost identical except for the fact that the larger simulation volume has a greater effective redshift and hence a power spectrum with lower amplitude and less non-linear evolution. We see that this holds true for our lightcone simulations, and that the difference in the replicated and unreplicated $512^{3}$ simulations is, at least on linear scales, equal to the difference in the linear growth factor between the effective redshifts of the two sets of simulations. 

However, in order to produce the replicated power spectrum, it is necessary to correct for the replication procedure. When we replicate a simulation, we are changing the fundamental mode of the simulation but without adding any additional information, either in the number of independent modes we sample, or on scales beyond the box size of the unreplicated simulation. This in turn creates ringing on the order of the unreplicated box size. This can demonstrated using a simple toy model. 

\begin{figure}
\centering
\includegraphics[width=84mm]{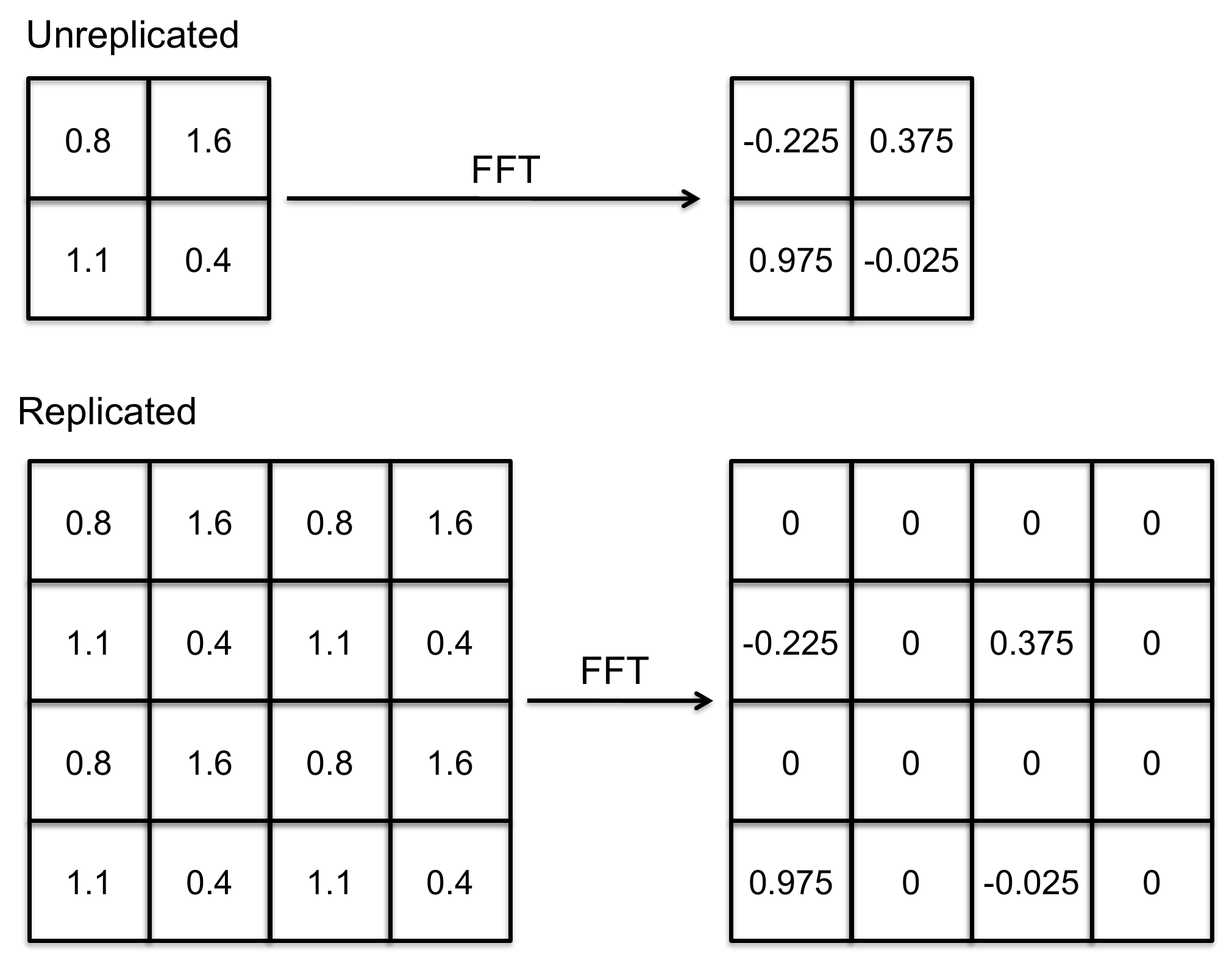}
\caption{A toy model demonstrating how replication of the simulation volume can create ringing in the power spectrum. Replicating the simulation does not add any information below the fundamental mode of the unreplicated simulation. The lack of additional information then creates '0' elements in the Fourier transformed overdensity grid, which in turn creates ringing in the power spectrum.}
\label{replicate_pk_toy}
\end{figure}

In Figure \ref{replicate_pk_toy} we show a small $2\times2$ overdensity field before and after taking the discrete Fourier transform. Then, if we replicate the $2\times2$ overdensity field 4 times and take the discrete Fourier transform, we assign the Fourier components to a grid 4 times larger than for the unreplicated field as the fundamental mode of the simulation should be twice as small. However we have not added any information beyond that contained in the original $2\times2$ grid and as such every other component of the Fourier transformed replicated field is zero, creating ringing within the power spectrum.

This also highlights the correction we perform to remove this affect. After Fourier transforming the replicated overdensity field we simply remove the zero components and place the remaining non-zero components in the same size grid as that used for the unreplicated box, correcting for the differences in normalisation between the two fields. We then compute the power spectrum using this smaller grid. This removes the ringing on the order of the box size and returns the power spectrum as seen in Figure \ref{replicate_pk}. It is important to note that this procedure still lacks the k-space resolution one would naively expect due to the fact our simulation box is larger. Neither our replication method nor our correction for ringing adds in modes larger than the unreplicated box size (there are, however, methods that do do this, see e.g. \citealt{Tormen1996, Cole1997})

This is an important correction and one that should be used whenever a simulation is replicated. It is important to note however that we believe such a correction to only be necessary when looking at a portion of a replicated simulation with volume equal to or greater than the unreplicated simulation. For most practical applications, the unreplicated simulation would be much larger than that used here, and the lightcone simulations themselves would undergo significant post-processing, such as the application of a survey window function and cutting into redshift slices. In this case the volume of each redshift slice will most likely be less than the original unreplicated simulation volume and so no correction will be necessary.

\subsubsection{Effects of Replication on the Covariance}

\begin{figure}
\centering
\includegraphics[width=84mm]{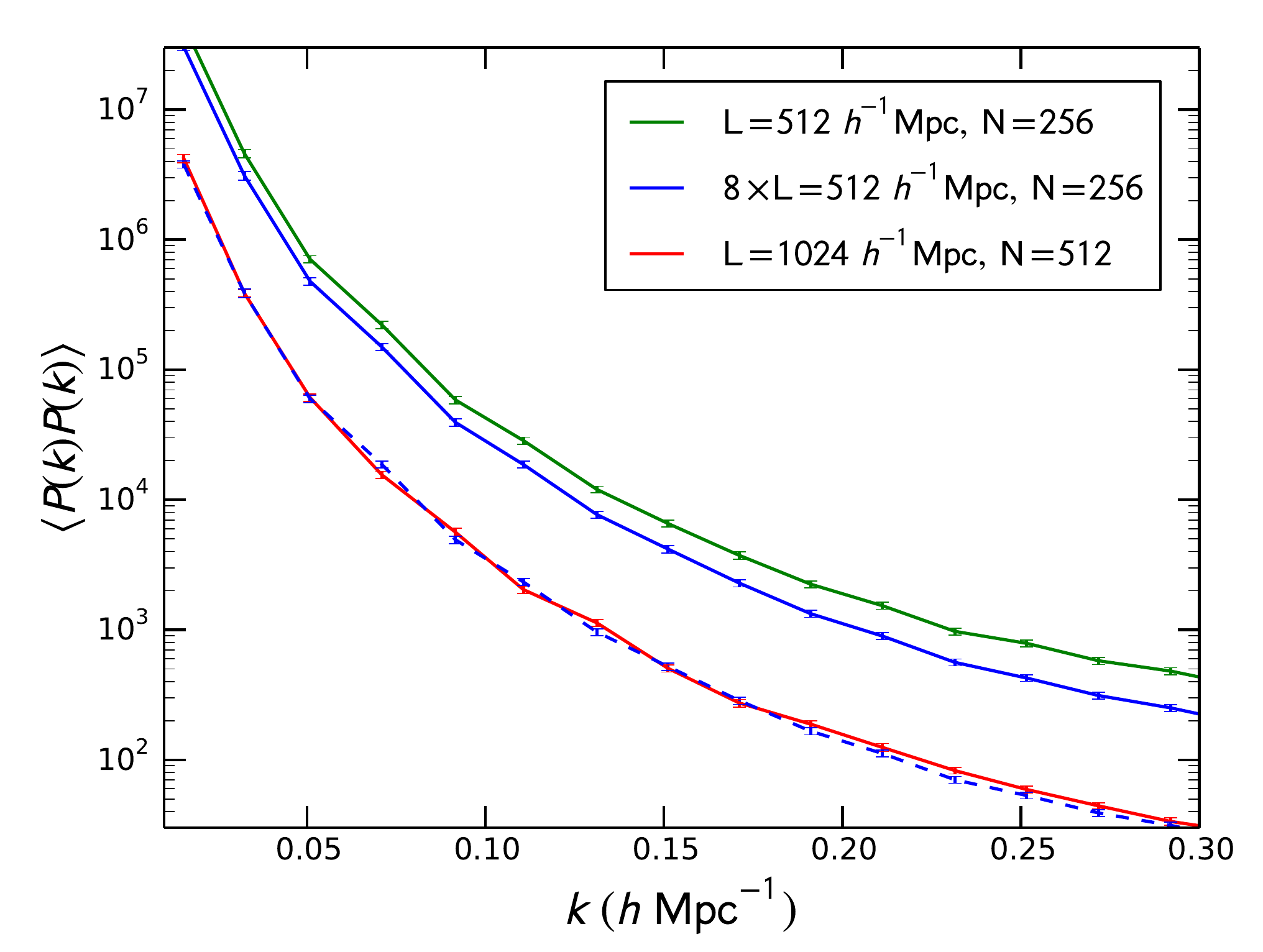}
\caption{Plots showing the effect of replication on the diagonal elements of the power spectrum covariance matrix using sets of $512^{3}$ particle lightcone simulations in a $1024\mpcoh$ box and $256^{3}$ particle lightcone simulations in a $512\mpcoh$ box. The errors are derived from bootstrap resampling with replacement over the 500 realisations. The dashed line shows the covariance of the replicated simulations after dividing by the difference in volume between the two sets of unreplicated simulations. The amplitude of the covariance matches the order of the legend.}
\label{replicate_cov}
\end{figure}

Utilising our 500 realisations for both sets of simulations we also look at the effect of replication on the covariance matrix. This is shown in  Figure \ref{replicate_cov}. Assuming Gaussian covariance, .i.e., \cite{Tegmark1997}, we would expect the covariance to scale as the inverse of the simulation volume. Our two sets of unreplicated simulations show this behaviour, with the larger volume simulation having a covariance 8 times smaller than the smaller simulation, at least on linear scales. But, as with the power spectrum, artificially increasing the simulation volume by replication does not add in any extra unique modes and so does not increase the variance. This in turn means that the covariance matrix of the replicated simulation does not display the expected volume dependence.

Knowing the expected volume dependence, however, we can correct for this effect. This correction is shown in \ref{replicate_cov} as the dashed blue line. The corrected, replicated covariance agrees very well with the unreplicated covariance, however there is some residual differences on small scales. We hypothesise that this arises due to the absence of modes larger than the unreplicated box size, which would otherwise couple with the small scale modes within the simulation and increase the small scale covariance. This coupling is referred to as the Super-Sample covariance by \cite{Takada2013} and \cite{Li2014}, who also explore corrections for this effect that could be applied to replicated simulations. 

However as, like the power spectrum, most applications of {\sc l-picola} will involve some manipulation of the final simulation output, we would not expect to see this incorrect volume dependence unless the comoving volume of the region we were analysing was close to the unreplicated simulation size. 

On the other hand, with this in mind, we still recommend that for any usage of {\sc l-picola} involving replication of the simulation region, the effects on the power spectrum and covariance matrix are throughly tested. This could be done using a procedure similar to that shown here, comparing replicated and unreplicated simulations after applying any survey geometry and data analysis effects. Obviously replication will only be necessary if maintaining both the full volume and number density is unfeasible, however as these effects arise due to the simulation volume rather than the particle number density one would be able to test this without simulating the full number of particles in the unreplicated volume.

\subsubsection{Speeding Up Replication}

In {\sc l-picola}, lightcone simulations are performed in such a way as to add no additional memory requirements to the run, however the amount of time to drift the particles will increase proportionally to the number of replicates. In order to speed this up we identify, each timestep, which replicates are necessary to loop over. Any replicates that have all 8 vertices inside the lightcone at the end of the timestep will not have particles leaving the lightcone and so can be ignored for the current iteration. Furthermore, for replicates not fully inside the lightcone, we calculate the shortest distance between the replicate and the origin by first calculating the distance to each face of the replicate then the shortest distance to each line segment on that face. If the shortest distance to the origin is larger than the lightcone radius then the replicate has completely exited the lightcone and will no longer be required for the duration of the simulation. Overall, this means that even if the simulation box is replicated $N$ times in each direction we will only need to look at a small fraction of the replicates ($\sim1-2$ in each direction unless the simulation box is so small that the lightcone radius changes by more than the boxsize in a single timestep).

\section{L-PICOLA Accuracy} \label{sec:accuracy}

\begin{figure*}
\centering
\subfloat{\includegraphics[width=0.45\textwidth]{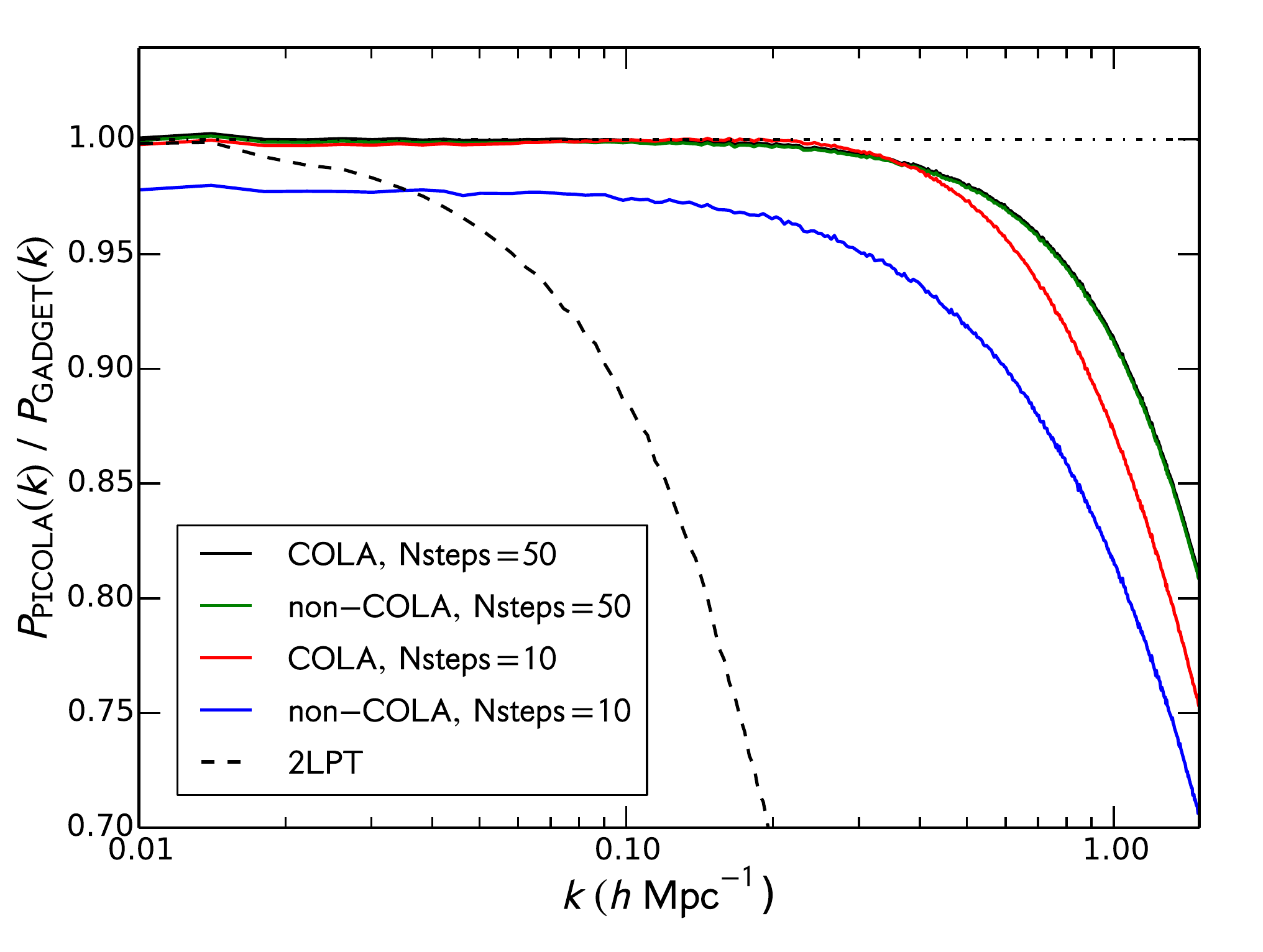}}
\subfloat{\includegraphics[width=0.45\textwidth]{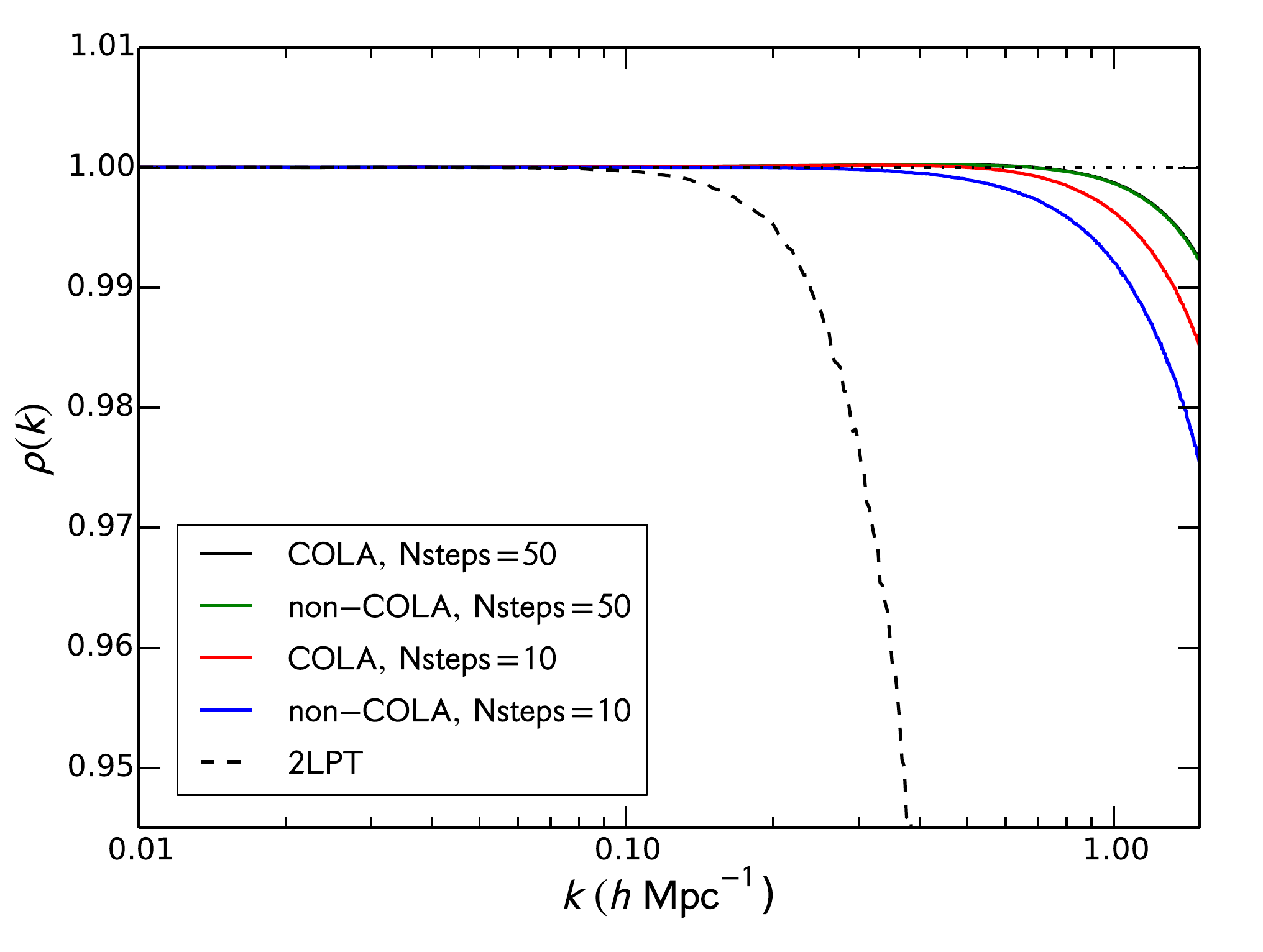}}
\caption{Plots of the power spectrum ratio and cross correlation between approximate realisations of the dark matter field, using the Particle-Mesh, 2LPT and COLA methods with 10 linear timesteps, and a Tree-PM realisation from {\sc gadget-2}. The amplitude of the ratios and cross correlations at large $k$ match the order of the legend.}
\label{picola_accuracy1}
\end{figure*}

In this section we compare the accuracy of {\sc l-picola} to a full N-Body simulation using the Tree-PM code {\sc gadget-2} \citep{Springel2005} and to the results returned using only 2LPT, which has been used to generate mock catalogues for the BOSS survey \citep{Manera2013, Manera2015}. In all cases we use a simulation containing $1024^{3}$ particles in a box of edge length $768\mpcoh$, starting at an initial redshift of 9.0 and evolving the dark matter particles to a final redshift of 0.0. We use our fiducial cosmology and a linear power spectrum calculated at redshift 0.0 from {\sc CAMB} \citep{Lewis2000, Howlett2012}. In all cases, unless this choice itself is being tested, we set $N_{mesh}=N_{particles}$. The memory requirements for this simulation are given in Appendix A, where this simulation is used as an example.

\subsection{Two-point Clustering}

We first look at how well {\sc l-picola} recovers the two-point clustering of the dark matter field compared to the N-Body simulation, which we treat as our fully correct solution. In all cases we estimate the power spectrum within our cubic simulations using the method of \cite{Feldman1994}. In Figure \ref{picola_accuracy1} we show the ratio of the power spectra recovered from the approximate simulations and from the {\sc gadget-2} run.  We plot the results recovered using 2LPT and {\sc l-picola} runs with 10 timesteps and 50 timesteps, and for a set of runs with the COLA modification turned off and the same numbers of timesteps. The act of turning the COLA method off reduces {\sc l-picola} to a standard Particle-Mesh code. We also plot the cross correlation, $\rho$, between the approximate dark matter field, $\delta$, and the full non-linear field from our N-Body run, $\delta_{NL}$, defined as
\begin{equation}
\rho(k) = \frac{\langle \delta(\boldsymbol{k})\delta^{*}_{NL}(\boldsymbol{k}) \rangle}{\langle|\delta(\boldsymbol{k})|^{2}\rangle \langle|\delta_{NL}(\boldsymbol{k})|^{2}\rangle}
\end{equation}
We neglect the contribution from shot-noise here as for dark matter particles and our simulation specifications this will be very sub-dominant on all scales we consider.

We see that the COLA method creates a much better approximation of the full non-linear dark matter field than 2LPT and the Particle-Mesh algorithms alone for a small number of timesteps. The agreement between the COLA and N-Body fields is remarkable, with the power spectra agreeing to within $2\%$ up to $k=0.3\hompc$, which covers all the scales currently used for BAO and RSD measurements. An $80\%$ agreement remains even up to scales of $k=1.0\hompc$. This level of conformity is mirrored in the cross correlation, which for the COLA run remains above 98\% for all scales plotted.

Further to this, where the cross-correlation is 1, we would not expect this to deviate between realisations. It is non-stochastic. As such we would expect that where the cross-correlation is 1, the covariance of the {\sc l-picola} and {\sc gadget-2} simulations would be identical (at the level of noise caused by using a finite number of realizations). Figure \ref{picola_accuracy1} indicates that the real-space covariance matrix recovered from {\sc l-picola}  is exact on all scales of interest to BAO and RSD measurements. Even where the cross-correlation between the {\sc l-picola} and {gadget-2} simulations  deviates from 1, it still remains very high, such that the covariance matrix recovered from {\sc l-picola} would match extremely well that from a full ensemble of N-Body realisations even up to $k=1.0\hompc$

In the same number of timesteps the Particle-Mesh algorithm cannot match the accuracy of COLA on any scales. Even on large scales there is a discrepancy between the PM and GADGET runs, as there are not enough timesteps for the PM algorithm to fully recover the linear growth factor. This validates the reasoning behind the COLA method as the 2LPT solution provides the solution on linear scales but performs much worse than the PM algorithm on smaller scales. The time taken for a single timestep under both the COLA and PM methods is identical and as such the COLA method gives much better results for a fixed computational time.

Interestingly, however, the COLA and the standard PM algorithm converge if a suitable number of timesteps is used (50 in this case). When this many timesteps are used the PM code can accurately recover the linear growth factor and the non-linear clustering is greatly improved. Using a larger number of timesteps for the COLA run only affects the non-linear scales as the linear and quasi-linear scales are already fully captured. Using larger and larger numbers of timesteps has a diminishing effect on both algorithms, as the small scale accuracy becomes bounded by the lack of force resolution below the mesh scale. As the COLA method is already quite accurate for a few timesteps increasing the number of timesteps for a fixed mesh size does not add as much accuracy as for the PM method alone. Incorporating the COLA mechanism into a Tree-PM code would negate this effect and we expect that increasing the number of timesteps used would then continue to increase the small scale accuracy beyond that achieved using the PM algorithm only. 

\begin{figure}[ht]
\centering
\includegraphics[width=84mm]{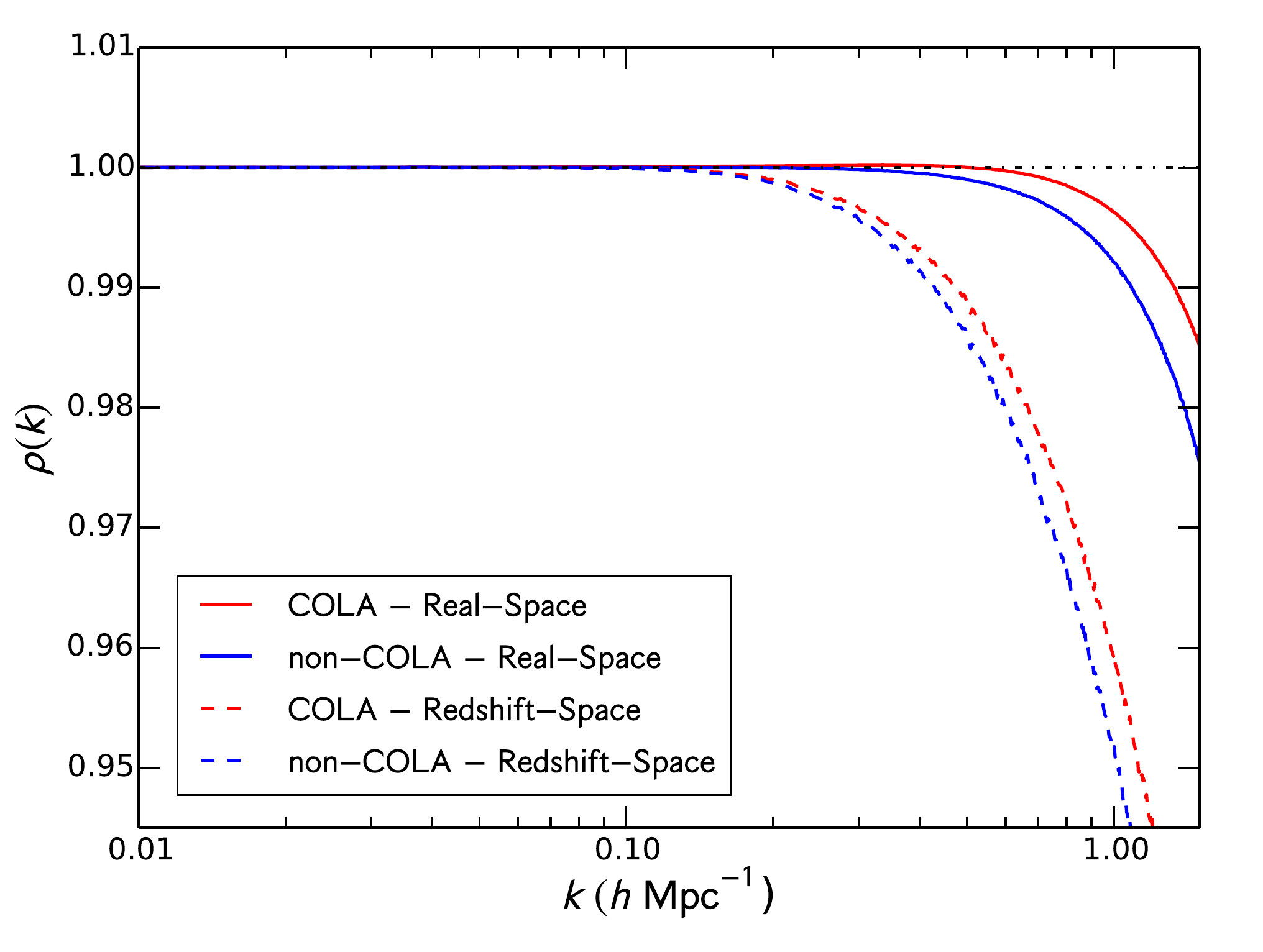}
  \caption{A comparison of the real- and redshift-space cross-correlations between approximate realisations of the dark matter field, using the Particle-Mesh and COLA methods with 10 linear timesteps, and a Tree-PM realisation from {\sc gadget-2}. The amplitude of the cross-correlation match the order of the legend.}
  \label{picola_accuracyrsd}
\end{figure}

Figure \ref{picola_accuracyrsd} compares the real and redshift-space cross correlation for the COLA and PM runs using 10 timesteps. The additional displacement each particle receives due to Redshift Space Distortions, $s_{los}$, is evaluated using
\begin{equation}
s_{los} = \frac{v_{los}}{H(a)a}
\end{equation}
where $v_{los}$ is the line of sight velocity of each particle for an observer situated in the centre of the simulation box.

In all cases we see that the accuracy of the simulation in redshift-space is worse than in real space. The $98\%$ cross correlation continues only up to $k=0.4\hompc$. However, this is to be expected as, in addition to slightly under-predicting the spatial clustering of the dark matter particles, the approximate methods do not recover the full non-linear evolution of the particle's velocities as a function of time. The agreement in redshift space between the COLA method and the {\sc gadget-2} run is still very good on all scales of interest to BAO and RSD measurements and the COLA method still outperforms the PM algorithm. Similarly we would expect the redshift-space covariance matrix to remain extremely accurate on these scales of interest.

\subsection{Three-point Clustering}
We also look at the accuracy with with {\sc l-picola} recovers the three-point clustering of the dark matter field. In particular we use the reduced bispectrum,
\begin{equation}
Q(k_{1},k_{2},k_{3}) = \frac{B(k_{1},k_{2},k_{3})}{P(k_{1})P(k_{2})+P(k_{2})P(k_{3})+P(k_{3})P(k_{1})}
\end{equation}
where $B(k_{1},k_{2},k_{3})$ is the bispectrum for our periodic, cubic simulation.

\begin{figure*}[ht]
\centering
\includegraphics[width=\textwidth]{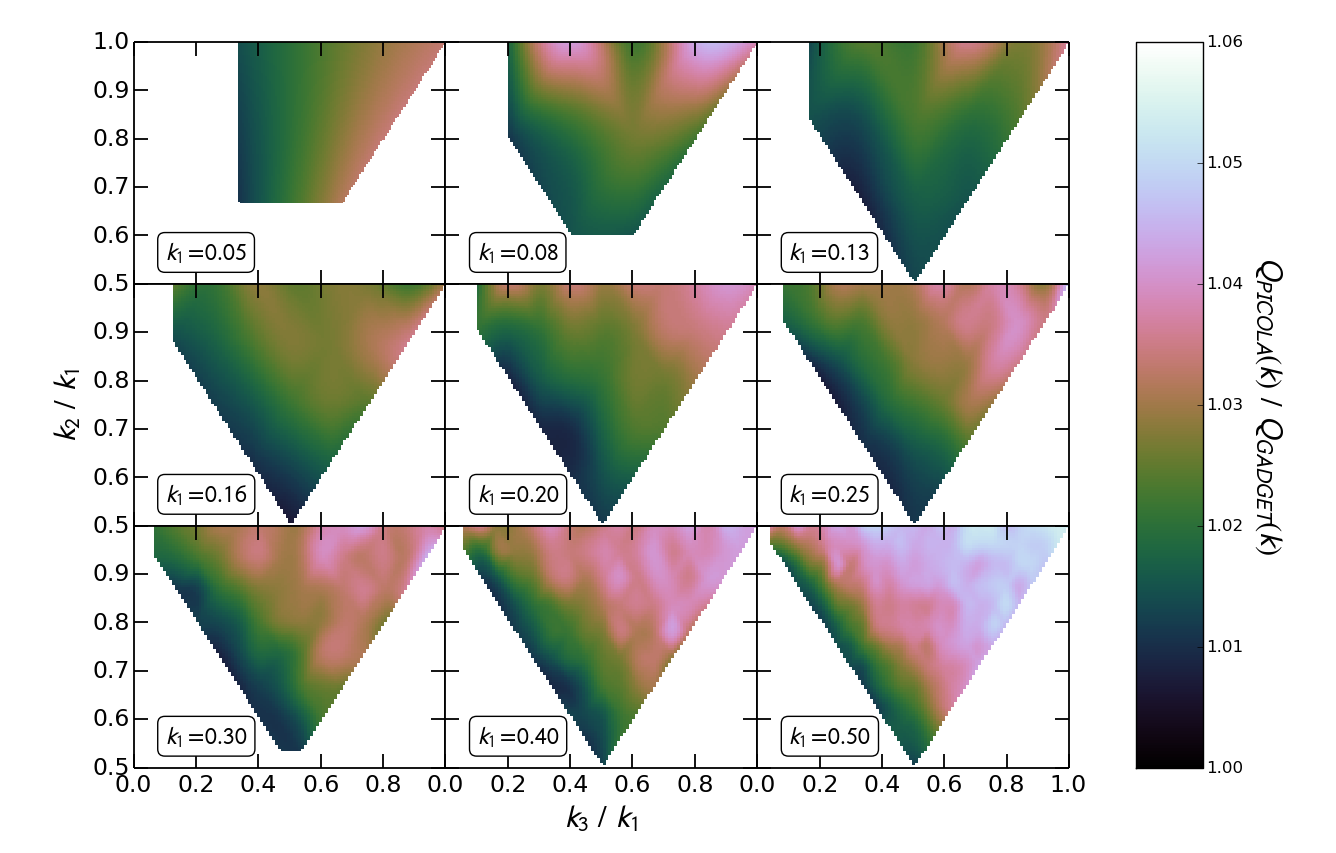}
  \caption{The ratio of the reduced bispectrum measured from our {\sc l-picola} and {\sc gadget-2} simulations. We plot this as a function of the ratios $k_{3}/k_{1}$ and $k_{2}/k_{1}$ for a range of $k_{1}$ values. This allows to use explore a wide range of triangle configurations. For reference the top-left, top-right and bottom vertices of each plot correspond to squeezed, equilateral and folded configurations, whilst the left,  and right and top edges correspond to elongated and isoceles triangles respectively.}
  \label{reduced_bispectrum2}
\end{figure*}

In order to explore the agreement between {\sc gadget-2} and {\sc l-picola} across a wide range of bispectrum configurations we plot the reduced bispectrum ratio for {\sc l-picola} and {\sc gadget-2} as a function of the ratios $k_{3}/k_{1}$ and $k_{2}/k_{1}$ for a variety of different values of $k_{1}$. This is shown in Figure \ref{reduced_bispectrum2}. For clarity in this figure and to avoid double plotting the same configurations we enforce the conditions $\vec{k_{1}}\ge \vec{k_{2}}\ge \vec{k_{3}}$ and $\vec{k_{1}}+\vec{k_{2}}+\vec{k_{3}}=0$.

From Figure \ref{reduced_bispectrum2} we find that {\sc l-picola} is able to reproduce the reduced bispectrum to within 6\% for \textit{any} bispectrum configuration up to $k_{1}=k_{2}=k_{3}=0.5\hompc$. We can also identify the configurations that {\sc l-picola} reproduces with greatest and least accuracy. Regardless of the scale we find that the bispectrum in the squeezed, elongated or folded limit is reproduced extremely well, to within $2\%$ on all scales. This is because these configurations contain large contributions from triangles with one or two large scale modes, which we expect {\sc l-picola} to reproduce exactly. The least accurate regime is the equilateral configuration, with accuracy decreasing as we go to smaller scales (larger $k_{1}$). This is because these triangles contain the biggest contribution from small scale modes in the simulation, which are not reproduced quite as accurately in {\sc l-picola}.

\subsection{Timestepping and Mesh Choices}

It should be noted that the convergence time of COLA depends intimately on the choice of timestepping and mesh size used and the accuracy after a given number of timesteps can vary based on the exact choices made. The representative run in Figure \ref{picola_accuracy1} uses the modified COLA timestepping and the value of $nLPT=-2.5$ suggested by \cite{Tassev2013} and a number of mesh cells equal to the number of particles.

\begin{figure}
\centering
\includegraphics[width=84mm]{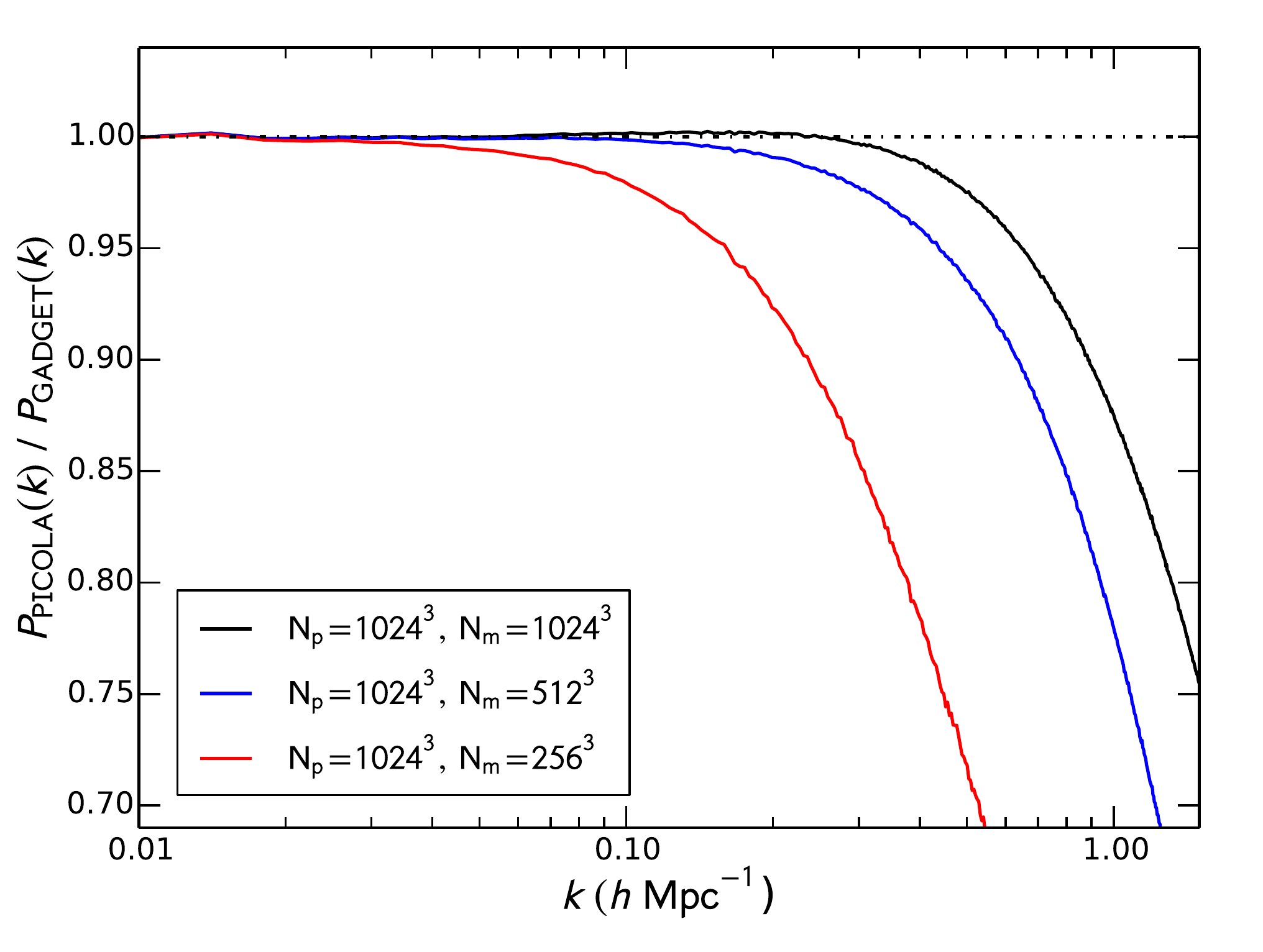}
  \caption{The power spectrum ratio between {\sc l-picola} dark matter fields using the COLA method and an N-Body realisation for different mesh to particle ratios. In all cases we run the simulation for 10 timesteps using linearly spaced COLA timesteps. The order of the legend matches the amplitude of the lines.}
  \label{picola_accuracymesh}
\end{figure}

Figure \ref{picola_accuracymesh} shows how the accuracy of COLA is reduced when lower force resolutions (less mesh cells) are used. We look at the case where the number of mesh points is equal to 1, 1/2 and 1/4 times the number of particles. We do not consider a number of mesh cells larger than the number of particles as, from the Nyquist-Shannon Sampling Theorem, we do not expect any improvement in the clustering at early times, when the particle distribution is approximately grid based. Furthermore \cite{Peebles1989} and \cite{Splinter1998} advocate that there is little justification in using a force resolution higher than the mean particle separation due to the inevitable differences in clustering between different simulations caused by using a finite number of particles. For most practical applications of {\sc l-picola} it also becomes computationally infeasible to use a number of mesh cells much larger than the number of particles, due to the large increase in computational time for the Fourier transforms.

As expected we find a reduction in the non-linear clustering accuracy as each mesh cell becomes larger, corresponding to a larger force smoothing. The large scales are still well recovered for all mesh sizes tested. Using smaller mesh sizes results in faster simulations and so for a given application of {\sc l-picola} a balance between mesh size and speed should be carefully considered based on the accuracy required and at which scales.

\begin{figure}
\centering
\includegraphics[width=84mm]{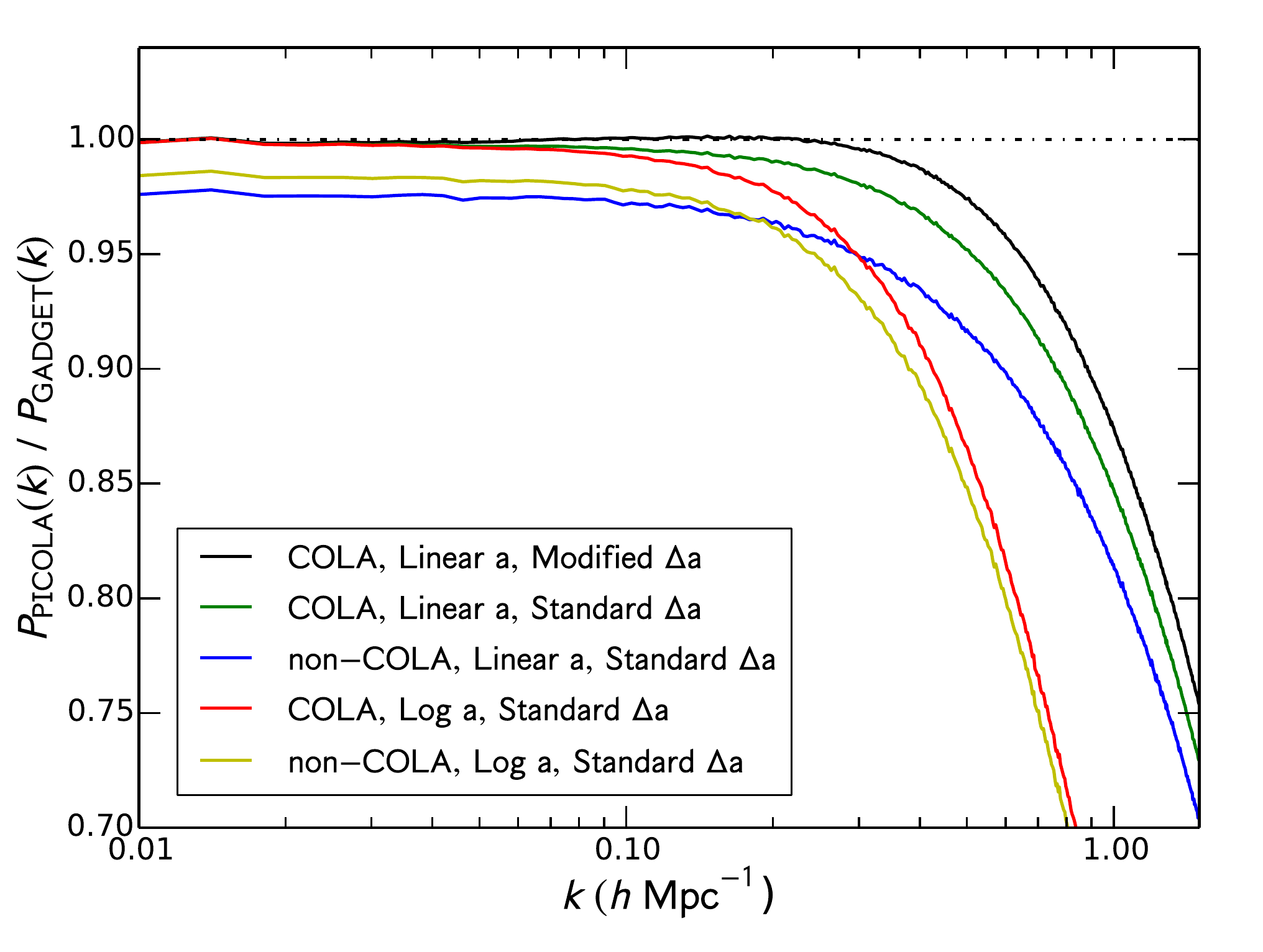}
  \caption{The power spectrum ratio between {\sc l-picola} dark matter fields and an N-Body realisation for different timestepping choices. We look at the effect of using timesteps linearly and logarithmically spaced in $a$ and using the modified method of \cite{Tassev2013} in place of the standard \cite{Quinn1997} timestepping. We also compare the COLA runs to standard PM runs using linearly and logarithmically spaced timesteps. In all cases we run the simulation for 10 timesteps. The order of the legend matches the amplitude of the lines at large $k$.}
  \label{picola_accuracy2}
\end{figure}

In Figure \ref{picola_accuracy2} we look at the effect of using timesteps linearly and logarithmically spaced in $a$ and also the effect of using the modified timestepping (with $nLPT=-2.5$ still) compared to the standard \cite{Quinn1997} method. 

In all cases we see that the COLA method still outperforms the standard Particle-Mesh algorithm, although to differing degrees. In the case of identical timestepping choices between the COLA and PM runs we see that the large scale and quasi-linear power is recovered much better. One point of interest is that using linearly spaced timesteps in the PM method reduces the accuracy on large scales below that of the logarithmically-spaced PM run, but greatly improves the non-linear accuracy, beyond even that of COLA with logarithmic steps. This is because using timesteps logarithmically spaced in $a$ means the code takes more timesteps at higher redshift, where the evolution of the dark matter field is more linear. This means that the PM algorithm recovers the linear growth factor more accurately. Using linear timesteps results in more `time' spend at low redshifts, where the evolution is non-linear and so the non-linear growth is captured more accurately, at the expense of the large scale clustering. As the COLA method gets the large scale clustering correct very quickly, using linear timesteps to increase the non-linear accuracy is much more beneficial. Indeed, we find even more improvement using the modified timestepping method, Eq.(\ref{eq:modifiedt}), which emphasises the non-linear modes and corroborates the claims of \cite{Tassev2013}.

\begin{figure*}
\centering
\subfloat{\includegraphics[width=0.45\textwidth]{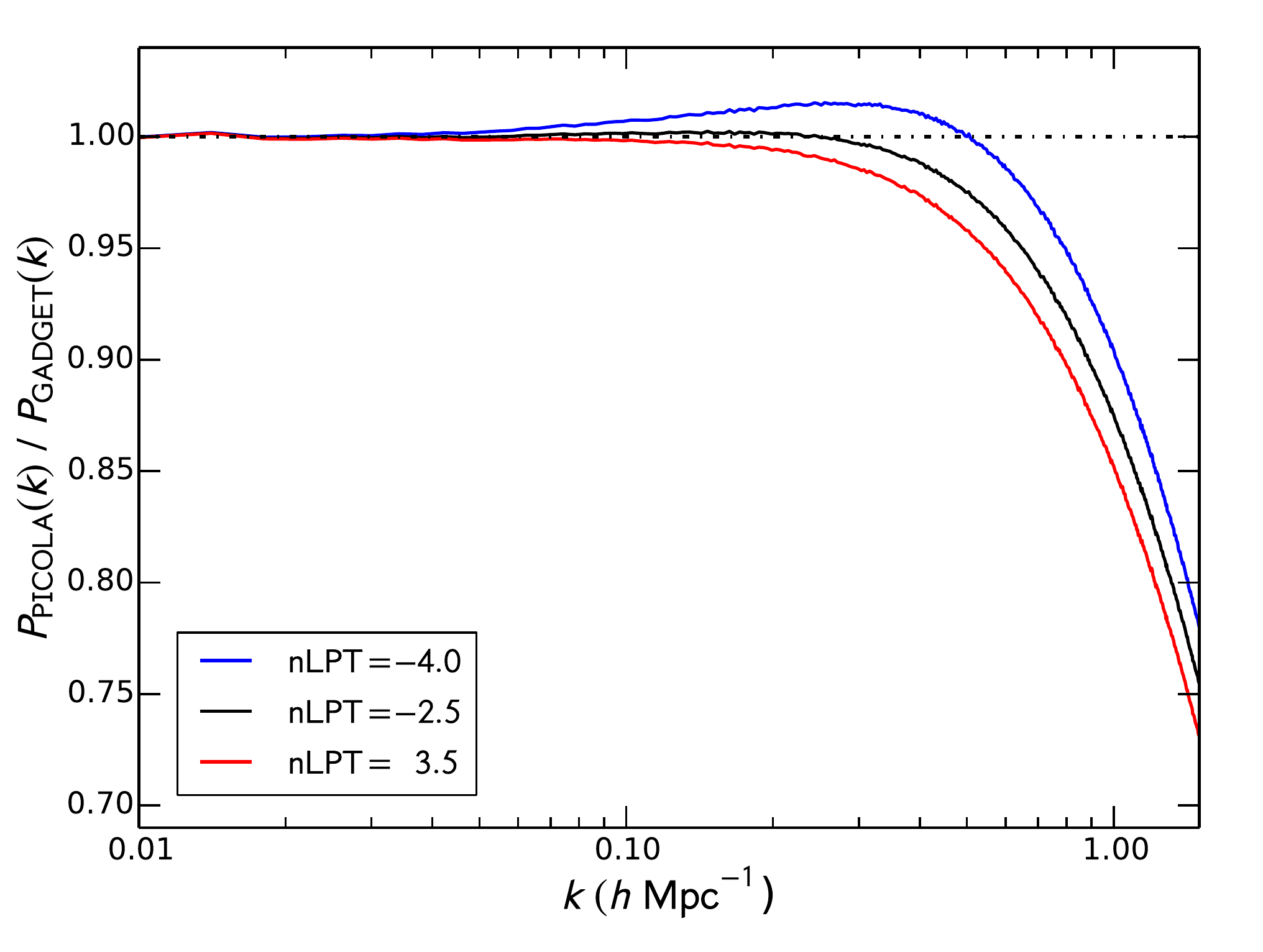}}
\subfloat{\includegraphics[width=0.45\textwidth]{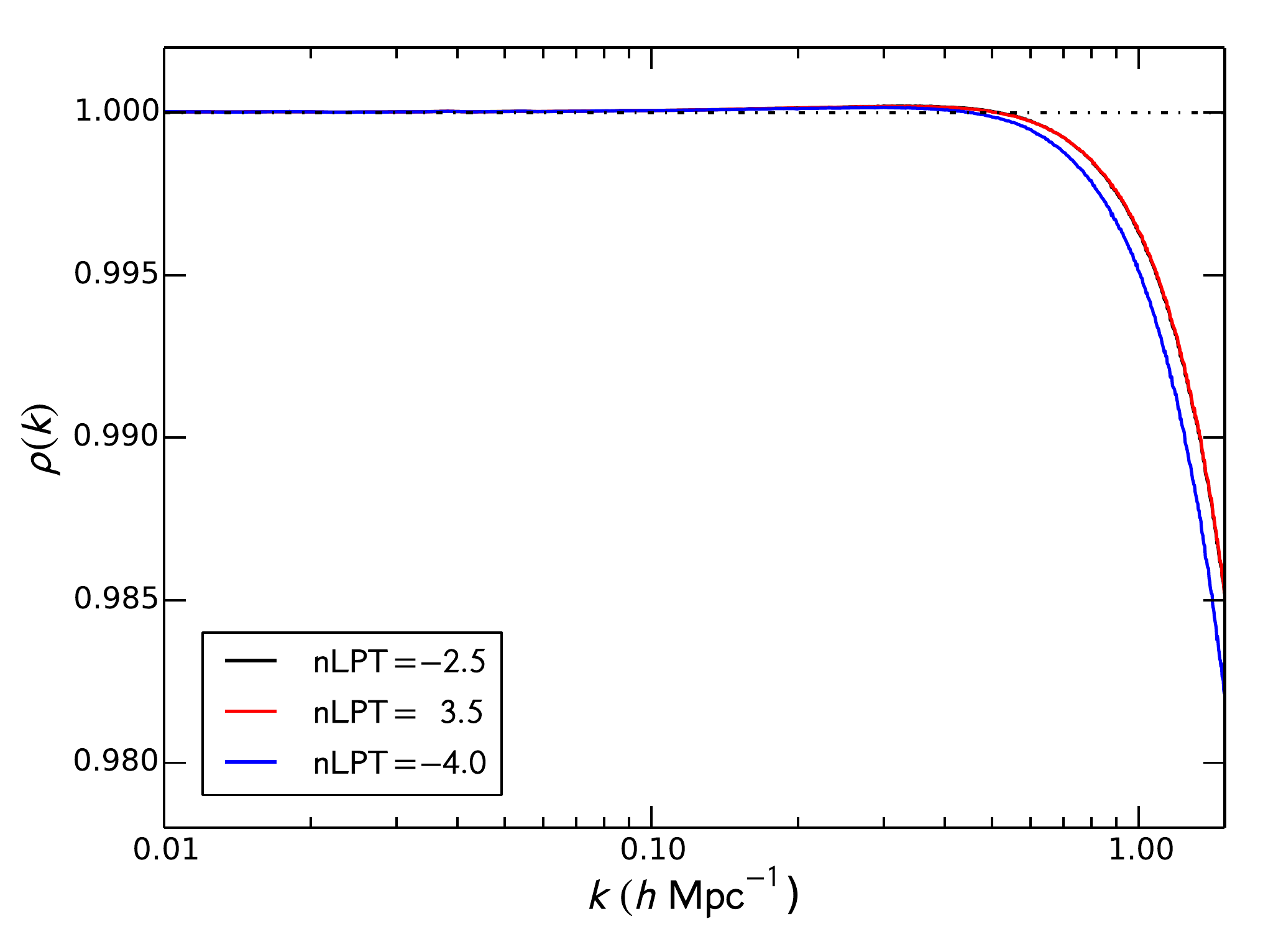}}
  \caption{The power spectrum ratio and cross-correlation between approximate dark matter fields made with {\sc l-picola} and an N-Body realisation for different values of $nLPT$ within the modifed COLA timestepping method. In all cases we run the simulations for 10 timesteps, using linearly spaced timesteps. The order of the legend matches the amplitude of the lines.}
  \label{picola_accuracy3}
\end{figure*}

It should be noted, however, that using the modified timestepping value puts additional emphasis on different growing modes, based on the value of $nLPT$, which can change the shape of the power spectrum. This is shown in Figure \ref{picola_accuracy3} where we plot the power spectrum ratio between the N-Body and {\sc l-picola} runs for different values of $nLPT$, exciting different combinations of decaying and growing modes, which are dominant at different cosmological times. We indeed see that different values produce slightly different power spectra. However the cross-correlations for these runs are all very similar, indicating that the difference is non-stochastic and cannot vary from realisation to realization.

As such, though the `correct' choice depends on the exact scales and statistics we wish to reproduce with our mock realisations, this is not very important. The results can be calibrated afterwards simply by comparing two different simulations with different values of $nLPT$. We find that for our case a value $nLPT=-2.5$ shows reasonable behaviour on all scales.

Throughout this section we have shown that the dark matter clustering recovered by {\sc l-picola} is extremely accurate on all scales of interest to BAO and RSD measurements. It is important to note however that when producing mock catalogues it is a representative galaxy field that is needed. In order to produce these {\sc l-picola} can be combined with other codes for identifying halos and populating the dark matter field with galaxies. Using the Friends-of-Friends algorithm \citep{Davis1985} and Halo Occupation Distribution model \citep{Berlind2002}, \cite{Howlett2015} generated mock catalogues from {\sc l-picola} fields. In this case no modification of the Friends-of-Friends linking length or the HOD model was needed. Other methods such as those presented by \cite{delaTorre2013}, \cite{Angulo2014} or \cite{Kitaura2014} could also be used.

\section{L-PICOLA Speed} \label{sec:speed}

We have shown that the COLA method itself outperforms both the 2LPT and Particle-Mesh algorithms in terms of the accuracy with which it reproduces the `true' clustering recovered from a Tree-PM N-Body simulations. In this section we highlight the transformation of the COLA method into a viable code for use with current and next generation large scale structure surveys by demonstrating the speed of {\sc l-picola} and showing how long it takes to produce a dark matter realisation compared to 2LPT and {\sc gadget-2}. 

We run a series of simulations with differing numbers of particles, box sizes and numbers of processors and look at the time taken in both the strong and weak scaling regimes. Strong scaling is defined as the change in the runtime of the code for different numbers of processors for a fixed simulation size, whereas weak scaling is the change in runtime for a fixed simulation size \textit{per processor}. For the strong scaling test we use the same simulation specifications as for our accuracy tests, with numbers of processors equal to \{8, 16, 32, 64, 128, 256\}. The simulations we use for the weak scaling are similar to those used for the strong scaling with additional details listed in Table \ref{tab:speedweak}. In all cases we fix the number of mesh cells to the number of particles.\footnote{All runs were performed on Intel Ivy Bridge CPU's on the {\sc SCIAMA} high-performance computing cluster at the University of Portsmouth. More information can be found at http://www.sciama.icg.port.ac.uk/}

\begin{table}
\centering
\caption{The specifications of the {\sc l-picola}, {\sc gadget-2} and 2LPT runs used in our weak scaling tests. In all cases we fix the number of mesh cells to the number of particles. All other simulation parameters are as used for the strong scaling runs and accuracy tests of Section \ref{sec:accuracy}.}
\begin{tabular}{ccc}
\hline
\hline
$N_{cpu}$ & $N_{particles}$ & $L_{box}$  ($\mpcoh$) \\[6pt]
\hline
$2$     & $256^{3}$   & $192 $\\
$4$     & $322^{3}$   & $246 $\\
$8$     & $406^{3}$   & $304 $\\
$16$   & $512^{3}$   & $384 $\\
$32$   & $644^{3}$   & $484 $\\
$64$   & $812^{3}$   & $610 $\\
$128$ & $1024^{3}$ & $768 $\\

\hline
\label{tab:speedweak}
\end{tabular}
\end{table}

\begin{figure*}
  \centering
  \addtocounter{subfigure}{-2}
  \subfloat[Full Simulation]{
  	\subfloat{\includegraphics[width=0.45\textwidth]{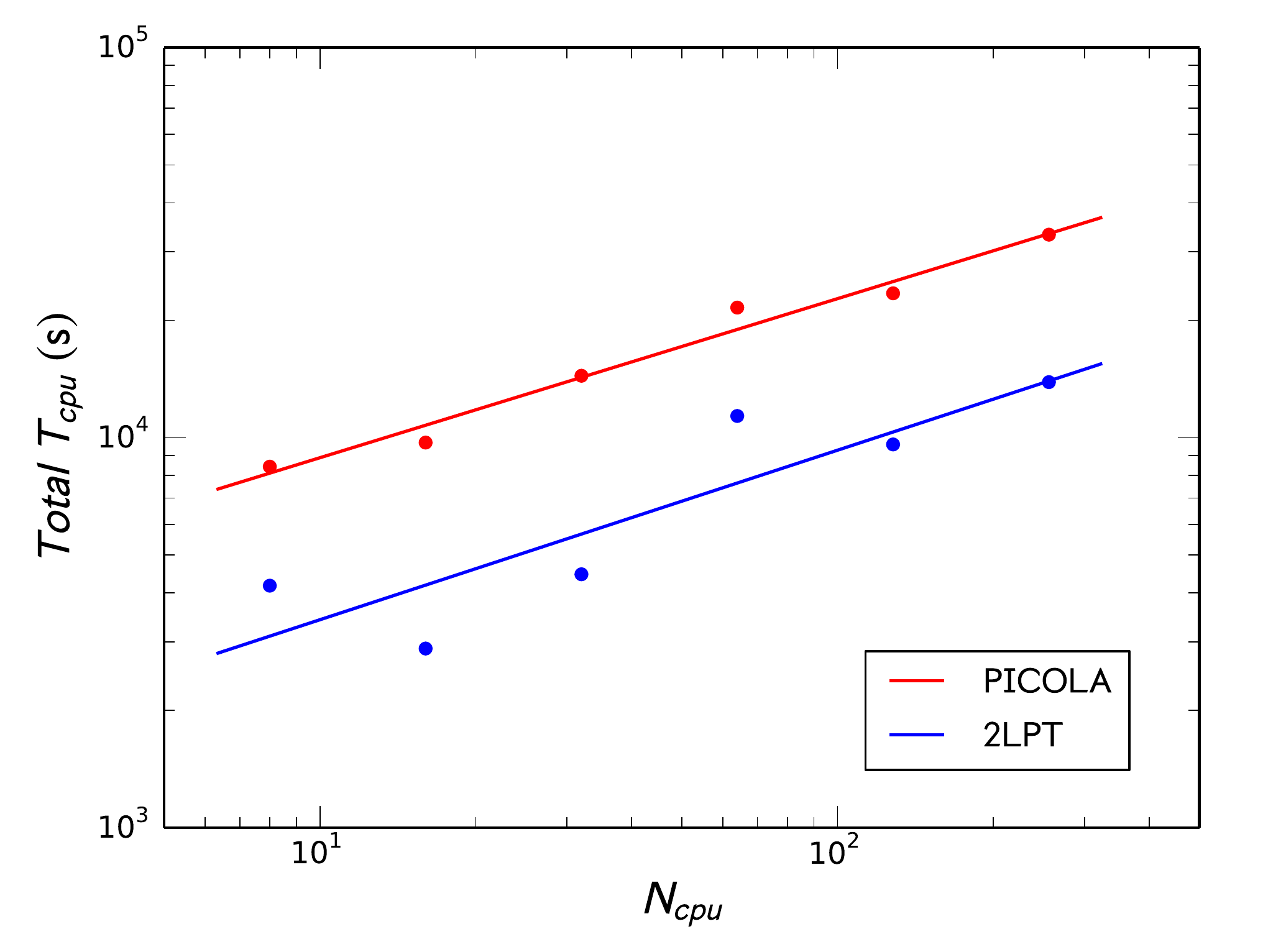}}
	\subfloat{\includegraphics[width=0.45\textwidth]{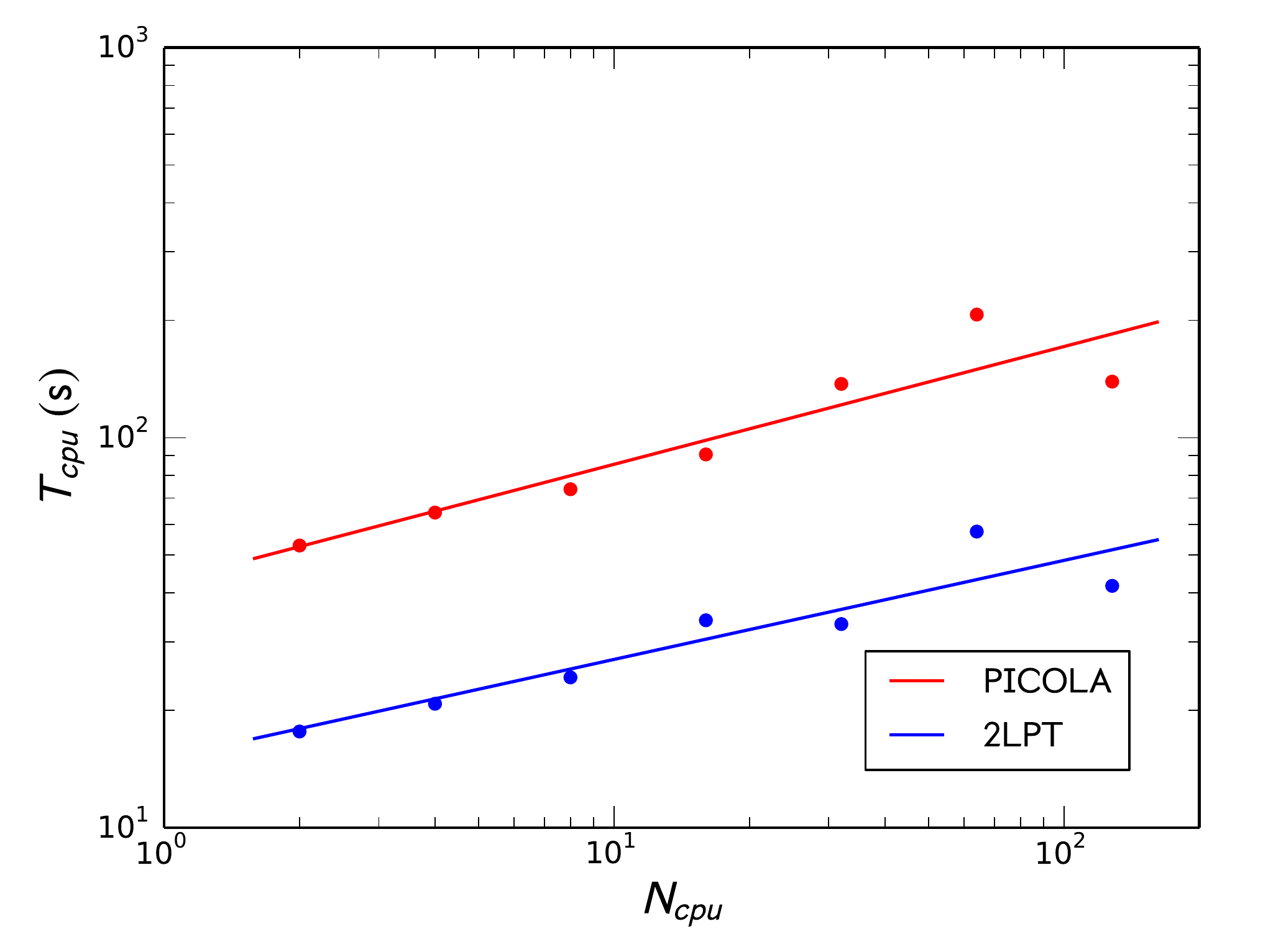}}} \\
	\addtocounter{subfigure}{-2}
  \subfloat[Simulation Contributions]{
  	\subfloat{\includegraphics[width=0.45\textwidth]{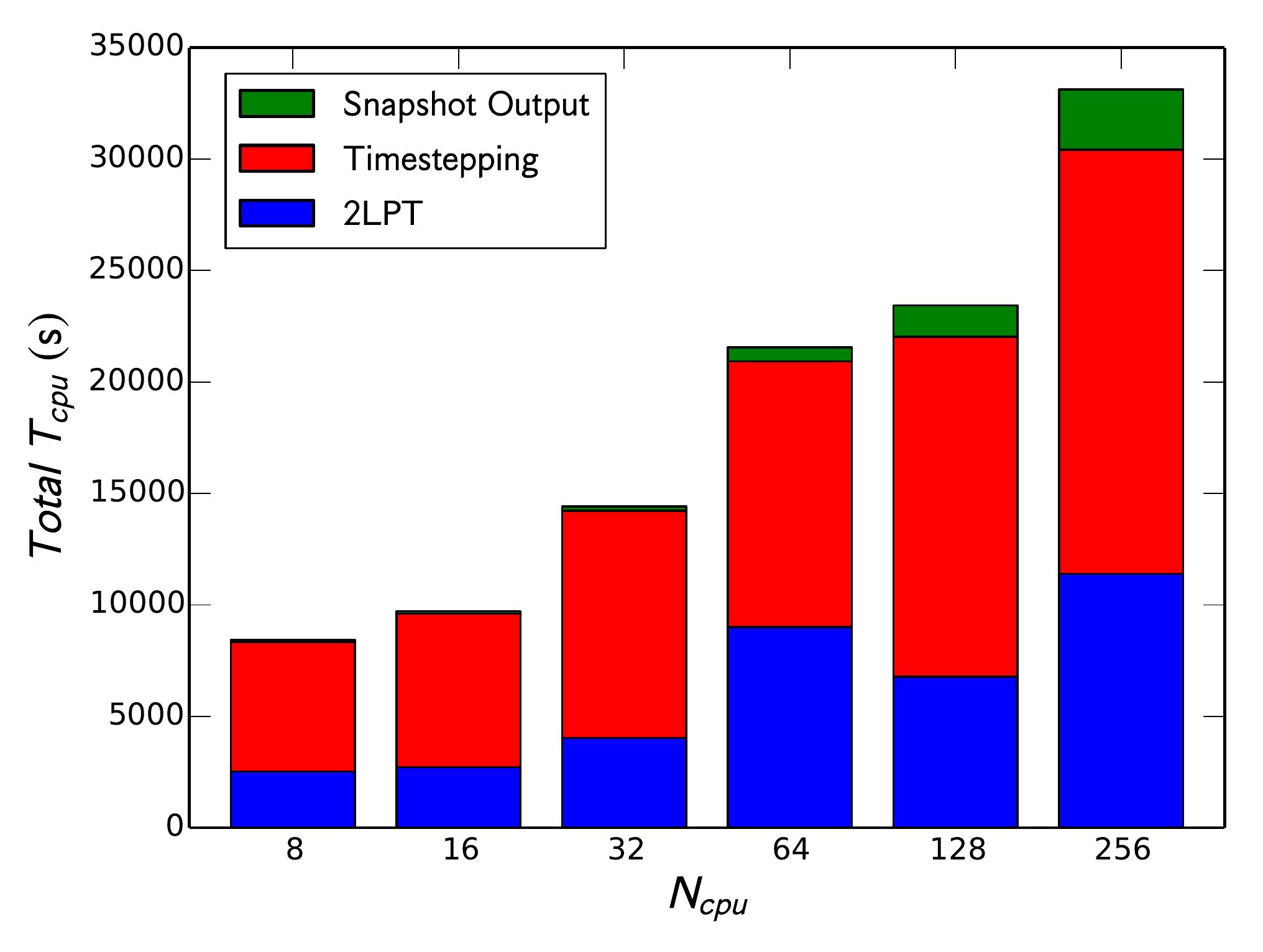}}
	\subfloat{\includegraphics[width=0.45\textwidth]{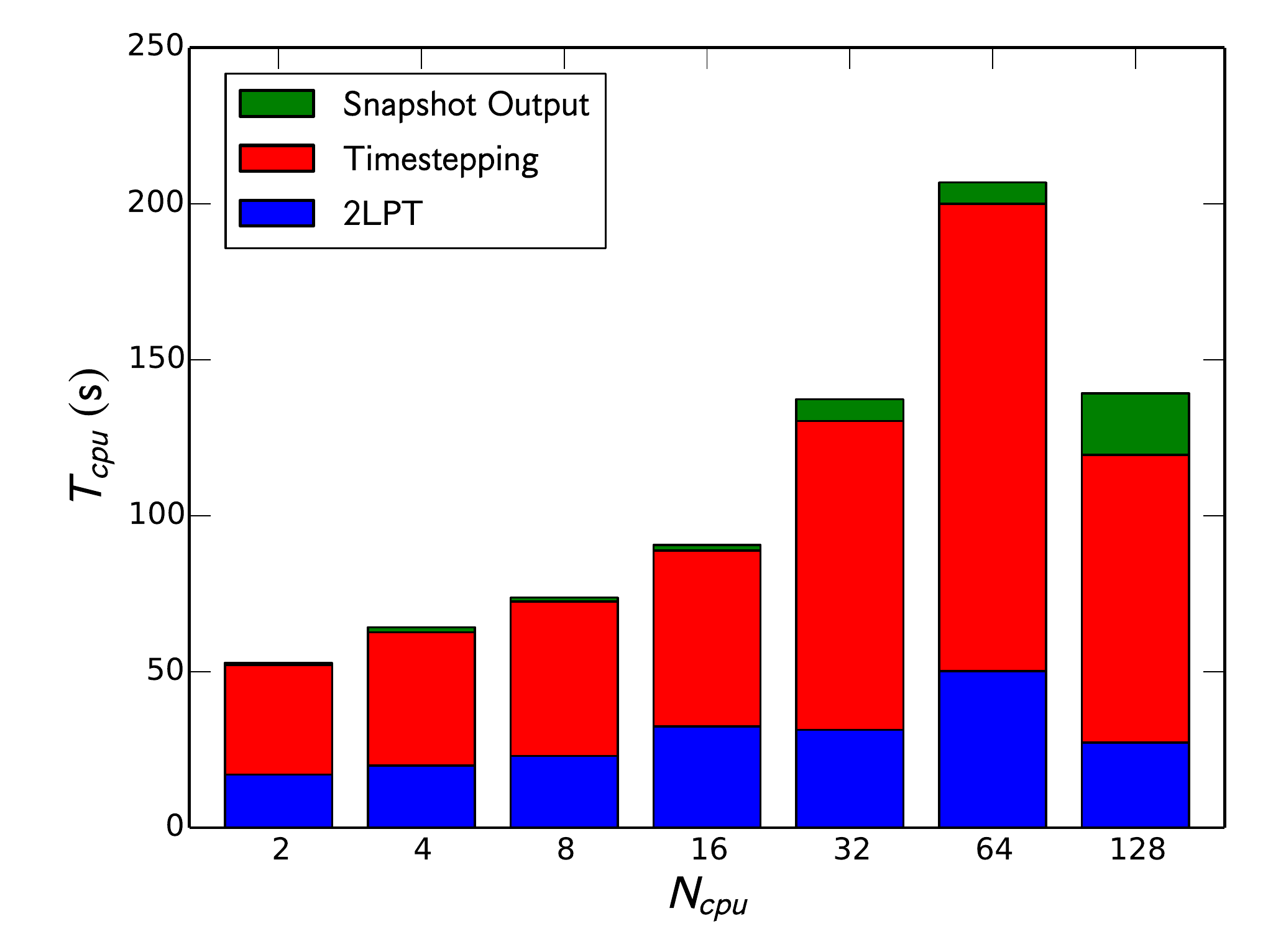}}} \\
	\addtocounter{subfigure}{-2}
  \subfloat[Timestep Contributions]{
  	\subfloat{\includegraphics[width=0.45\textwidth]{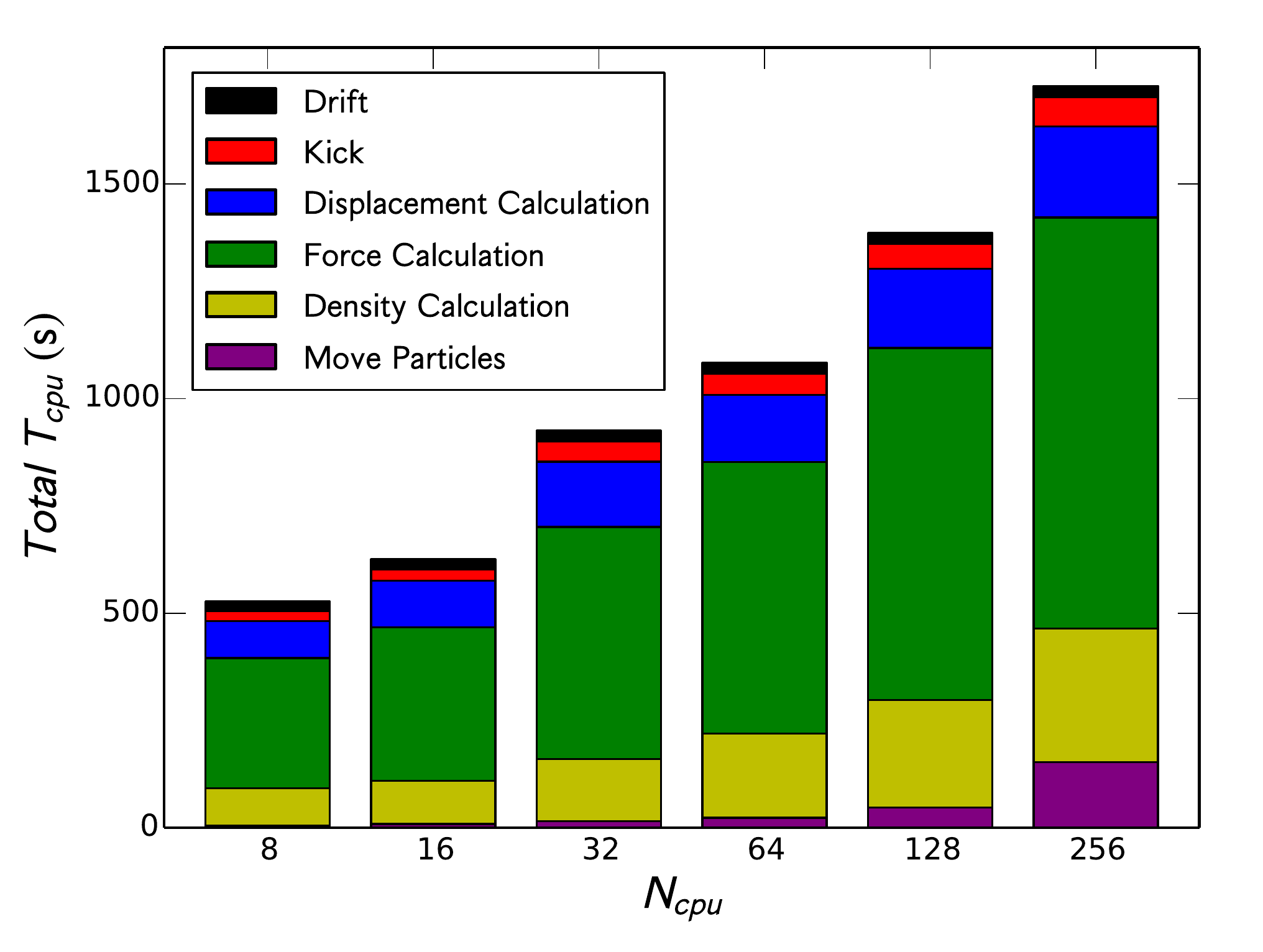}}
	\subfloat{\includegraphics[width=0.45\textwidth]{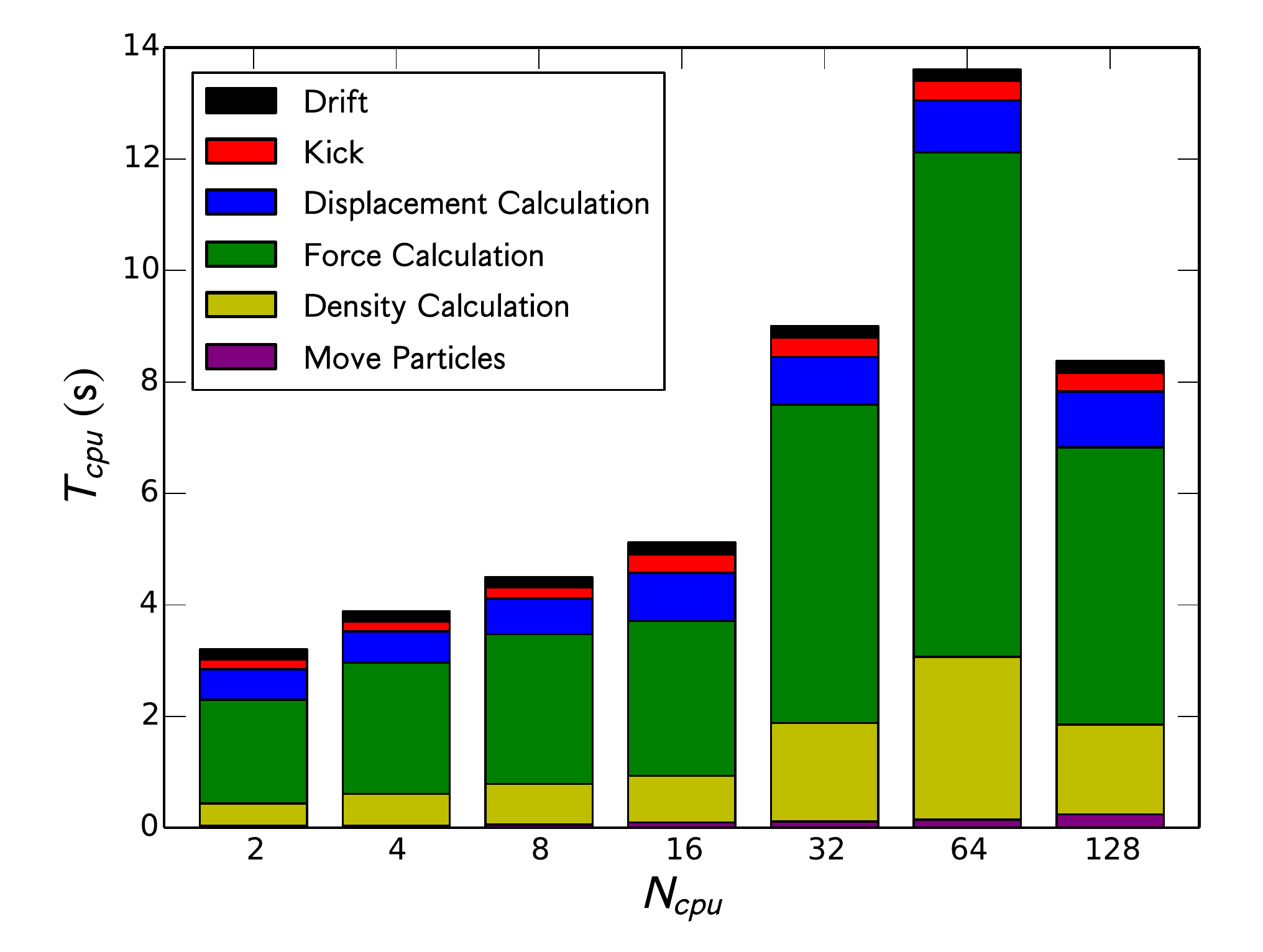}}} \\
  \caption{Plots showing the scaling of {\sc l-picola} in the strong (left) and weak (right) scaling regimes. For the strong scaling we plot the total CPU time, summed across all processors, whilst for the weak scaling we plot the CPU per processors. This means that ideal scaling would be shown as a constant horizontal trend as a function of the number of processors. Different panels show the total time taken for {\sc l-picola} compared to 2LPT simulations; the different contributions to the {\sc l-picola} runtime; and the contributions to a single {\sc l-picola} timestep. In all cases the order of the legend matches the order of the plotted quantities.}
  \label{picola_scaling}
\end{figure*}

All the run times are shown in Figure \ref{picola_scaling}, for both the strong and weak scaling. In both cases we have plotted the CPU time in such a way that perfect scaling will result in a constant horizontal line (total CPU time summed across all processors for strong scaling and CPU time per processor for weak scaling). The top panel of this Figure shows the full CPU time taken for each run. We find that our {\sc l-picola} runs generally take about 3 times longer to run that a simple 2LPT realisation, however this is a relatively small cost compared to the difference in the accuracy of the methods.

In terms of the actual scaling we find that although {\sc l-picola} does not scale perfectly in either the strong or weak regimes, the increase in runtime with number of processors is still reasonable. We use a simple least squares fitting method to fit a linear trend to the CPU time as a function of the number of processors. We find gradients of 0.41 and 0.30, compared to the ideal value of 0, for {\sc l-picola} in the strong and weak scaling regimes respectively. We find that this trend can be extrapolated well beyond our fitting range., i.e., for a $2048^{3}$ particle simulation in a ($1536\mpcoh)^{3}$ box run on 1024 processors we find a CPU time per processor of 348 seconds, which matches very well our predicted value of 345 seconds.

There exists some scatter in the runtimes for our simulations. This generally stems from the Fourier transforms involved, the efficiency of which depends on the way the mesh in partitioned across the processors. In the case where we have a number of mesh cells (in the x-direction) that is not a multiple of the number of processors (.i.e, the $N_{cpu}=64$, strong-scaling run) the time taken for the calculation of the 2LPT displacements and the interparticle forces during timestepping is increased.  

\subsection{Contributions to the runtime}

We investigate this further in the second and third rows of Figure \ref{picola_scaling}. Here we show the time taken for different contributions to the full run and to each timestep therein. This highlights the fact that the scatter occurs mainly during the 2LPT and force calculation parts of {\sc l-picola} as expected if it is due to the Fourier transform efficiency. Additionally this also suggests that the non-optimal strong and weak scaling does not stem from any particular part of the code, but rather due to the extra MPI communications needed when we use larger numbers of processors.

Looking at the contributions from the 2LPT, Timestepping and Output stages, we see that there is some evolution with processor number in the 2LPT stage from the fact that the Fourier transforms require extra communications between different processors to transform the full mesh. We also see an increasing contribution from the Output stage of the code as we go to larger numbers of processors. This is because of an option in the code to limit the number of processors outputting at once, stopping all processors outputting simultaneously. We set this to 32 processors and as expected we see an increase in the time taken to output the data once we run simulations with more than 32 processors due to the need for some processors to wait before they can output.

\subsection{Contributions to a single timestep}

Looking at the contributions to an individual timestep we find that the Drift, Kick and Displacement parts of the code are reasonably constant when the number of particles per processor remains constant. These consist mainly of loops over each particle and so this is too be expected. The Density Calculation and Force Calculation steps contain the Fourier transforms required for each timestep and as such are the biggest contributions to the time taken for a timestep. Looking at the strong scaling case we see an increase in both of these as a function of the number of processors, which indicates they are dominated by the MPI communications as expected. 

For the weak scaling we also see a large jump in the CPU time for both of these after 16 processors. This is another indication that the MPI communications are the cause of the scaling trends we see, as the architecture of the High Performance Computer we use is such that 16 processors are located on a single node and intra-node communication is much faster than inter-node. Once we start to require inter-node communication to compute the Fourier transforms the CPU time increases.

Finally we see that the Move Particles section of the code does not contribute much to the total time for each timestep, except where the number of inter-processor communications becomes large. This is due to the effort taken to produce a fast algorithm to pass the particles, whereas a simpler algorithm would result in a larger amount of time and memory needed to identify and store the particles that need transferring.

\section{Conclusion}
In this paper we have introduced and tested a new code {\sc l-picola}, which, due to its fast nature, can be used to generate large numbers of accurate dark matter matter simulations. The code is available under the GNU General Public License at \url{https:/cullanhowlett.github.io/l-picola}. The main points of the paper are summarised as follows:
\begin{itemize}
\item{{\sc l-picola} is a memory conservative, planar parallelisation of the 2LPT and COLA algorithms. This is enabled by parallel algorithms for Cloud-in-Cell interpolation, Fast Fourier Transforms, and fast movement of particles between processors after each timestep.}
\item{We have included additional features in {\sc l-picola} such as the fast creation of initial conditions for other simulation codes, with optional primordial non-Gaussianity, and the ability to produce lightcone simulation, with optional replication of the simulation volume at run-time. These will be of particular use to future large scale structure surveys.}
\item{We have quantified the accuracy of the method {\sc l-picola} uses to produce lightcone simulations, verifying that it's accuracy is not unduly affected by the approximations made to ensure a fast algorithm.}
\item{We have investigated the effect of replicating the simulation volume on both the power spectrum and covariance matrix using a set of 500 individual lightcone realisations. We find that, due to the fact the replication procedure modifies the simulation volume without adding additional information, the power spectrum can suffer from ringing on the scale of the unreplicated box size and that the covariance matrix demonstrates the volume dependence of the unreplicated box size as opposed to the replicated volume. We show simple corrections for both of these effects and hypothesise that this is only a problem when analysing regions of the simulation larger than the unreplicated box size.}
\item{We have compared the accuracy of {\sc l-picola} to the approximate 2LPT and PM methods and to a fully non-linear Tree-PM {\sc gadget-2} simulation. We find that {\sc l-picola} performs much better than the 2LPT and PM algorithms, and that the power spectra from {\sc l-picola} agree with that from our {\sc gadget-2} simulations to within $2\%$ on all scales of interest to BAO and RSD measurements and to within $20\%$ up to $k=1.0\hompc$. The reduced bispectrum from {\sc l-picola} also shows remarkable agreement with our {\sc gadget-2} simulation, to within 6\% for all configurations up to $k_{1}=k_{2}=k_{3}=0.5\hompc$. We do however find that this agreement has some dependence on the exact type of timestepping used in the code.}
\item{We have compared the speed of {\sc l-picola} to the 2LPT and {\sc gadget-2} simulations. We find that the remarkable accuracy of {\sc l-picola} comes at only a small cost to speed compared to 2LPT. {\sc l-picola} exhibits reasonable scaling properties in the strong and weak scaling regimes, even up to large numbers of processors. We find that these trends are dominated by the need for extra inter-processor communication when using large numbers of processors.}
\end{itemize}

Still, there are several improvements that could be made to {\sc l-picola} in the future. In terms of parallelisation, splitting the mesh into 'blocks'
rather than 'slices' could improve both the speed and scalability of the code to large numbers of processors, however the need for additional MPI communication during the Fast Fourier Transforms means that the level of improvement is indeterminate at this time. Furthermore one could imagine hybridising the code, using Open-MP and MPI such that communication between `local' processors does not rely on slower MPI communication.

In terms of the physics behind {\sc l-picola} it would be simple to add in support for warm dark matter. Another obvious addition to the code would be to implement the spatial extension of the COLA method, presented by \cite{Tassev2015}. Such an improvement would allow us to simulate a large cosmological volume whilst only spending computational time evaluating the non-linear displacements for a small portion of that volume. Lightcone simulations within {\sc l-picola} in particular would greatly benefit from this as we would be able to simulate a small 'pencil-beam' region of the full lightcone and scale this up to the required simulation volume. Also, as this extended COLA method still requires us to calculate the 2LPT displacements for all the particles within the full volume, implementing this into our distributed-memory code would allow us to simulate much larger cosmological volumes and higher particle densities than the current shared-memory implementation.

Additional small scale accuracy could be achieved by a suitable scaling of the mesh during the simulation, such as using a finer mesh at late times when the particles become more clustered. This would be particularly easy to implement as, in the optimal memory case, the mesh is deallocated and reallocated each time step anyway. Using an adaptive mesh for high density portions of the simulation, or the Tree-PM algorithm instead of the PM algorithm, could also be implemented though these methods would come at a cost to speed.

Furthermore, as {\sc l-picola} is so fast, we find that for current applications, the total CPU time taken to produce a mock galaxy catalogue is dominated by outputting and post-processing of (mainly reading in) the dark matter field, especially the creation of dark matter halos. This is exacerbated even more for lightcone simulations with replication as we are effectively outputting the simulation multiple times, resulting in large increases to the amount of time taken to output and process the data. This could be vastly improved by adding in a halo finder into {\sc l-picola}, either by identifying shell-crossing as it occurs during the simulation, or via the FoF algorithm. This would mean that the amount of time taken to output the data, and read it in for post-processing could be reduced drastically.

\section*{Acknowledgements}

We would like to thank Angela Burden, Gary Burton, James Briggs and John Pennycook for their help and insightful comments throughout the development and testing of {\sc l-picola}.

We make special acknowledgement to the facilities and staff of the UK Sciama High Performance Computing cluster supported by the ICG, SEPNet and the University of Portsmouth. Code development and testing, mock catalogue generation, power spectrum estimation, and other analysis all made use of the computing and storage offered therein. This research would not have been possible without this support.

CH is grateful for funding from the United Kingdom Science \& Technology Facilities Council (UK STFC) grant \\ST/K502248/1. WJP acknowledges support from the UK STFC through the consolidated grant ST/K0090X/1, and from the European Research Council through grants 202686 (MDEPUGS) and 614030 ({\it Darksurvey}).

This work used the DiRAC Data Analytic system at the University of Cambridge, operated by the University of Cambridge High Performance Computing Service and the COSMA Data Centric system at Durham University, operated by the Institute for Computational Cosmology, both on behalf of the STFC DiRAC HPC Facility (www.dirac.ac.uk). This equipment was funded by BIS National E-infrastructure capital grants \\ ST/K001590/1 and ST/K00042X/1, STFC capital grants \\ ST/H008861/1 and ST/H00887X/1, and STFC DiRAC Operations grants ST/K00333X/1 and ST/K00087X/1. DiRAC is part of the National E-Infrastructure.

This work was also undertaken on the COSMOS Shared Memory system at DAMTP, University of Cambridge operated on behalf of the STFC DiRAC HPC Facility. This equipment is funded by BIS National E-infrastructure capital grant \\ ST/J005673/1 and STFC grants ST/H008586/1, ST/K00333X/1.

This research has made use of NASA's Astrophysics Data System Bibliographic Services.

\appendix
\section{Memory Consumption}
Considerable effort has been made to reduce the memory footprint of {\sc l-picola} as much as possible, including the introduction of a compilation option to conserve as much memory as possible. When this option is used the memory consumption for a {\sc l-picola} run is reduced significantly and the mean memory per processor can be calculated reasonably simply.

Here we detail the calculation of the memory needed for an {\sc l-picola} simulation under these optimum conditions. With the optimal memory setting we use floating point precision for the particles and double precision for the mesh. The information for each particle consists of x, y and z coordinates, velocities in those same directions, and the ZA and 2LPT displacements in those directions, resulting in $M_{p}=48\mathrm{Bytes}$ per particle. The main contributions to the memory arise from the particles and the mesh and the key parameters are the number of mesh cells, $N_{m}$, number of particles, $N_{p}$ and the amount of buffer memory allocated to each processor to account for the non-uniformity of the particle distribution over processors at late times, $b$.

The code can be split into six distinct sections: the calculation of the initial 2LPT potentials; the calculation of the initial 2LPT displacements; the initialisation of the particles; the moving of particles across processors each timestep; the evaluation of the interparticle mesh-based force each timestep; and the calculation of the particle displacements for each timestep. The corresponding memory requirements are:
\begin{equation}
M_{2LPT} = \frac{72N_{m}^{2}(N_{m}+2)}{N_{proc}} + 72N_{m}(N_{m}+2) + 4N_{m}^{2}
\end{equation}
\begin{equation}
M_{DISP}  = \frac{48N_{m}^{2}(N_{m}+2)+24N_{p}^{3}}{N_{proc}} + 48N_{m}(N_{m}+2)
\end{equation}
\begin{equation}
M_{INIT} = \frac{(24+bM_{p})N_{p}^{3}}{N_{proc}}
\end{equation}
\begin{equation}
M_{MOVE} = \frac{(b+2(b-1))M_{p}N_{p}^{3}}{N_{proc}}
\end{equation}
\begin{equation}
M_{DENS} = \frac{32N_{m}^{2}(N_{m}+2)+bM_{p}N_{p}^{3}}{N_{proc}} + 40N_{m}(N_{m}+2)
\end{equation}
\begin{multline}
M_{NBODY} = \frac{(12+bM_{p})N_{p}^{3}+24N_{m}^{2}(N_{m}+2)}{N_{proc}} \\
+ 24N_{m}(N_{m}+2)
\end{multline}
The maximum memory required for an {\sc l-picola} simulation is the largest of these 6 contributions. A utility for calculating the memory requirements, even when using suboptimal (in terms of memory) compilation options is provided with the public release of the code.

\begin{figure}
\centering
\includegraphics[width=84mm]{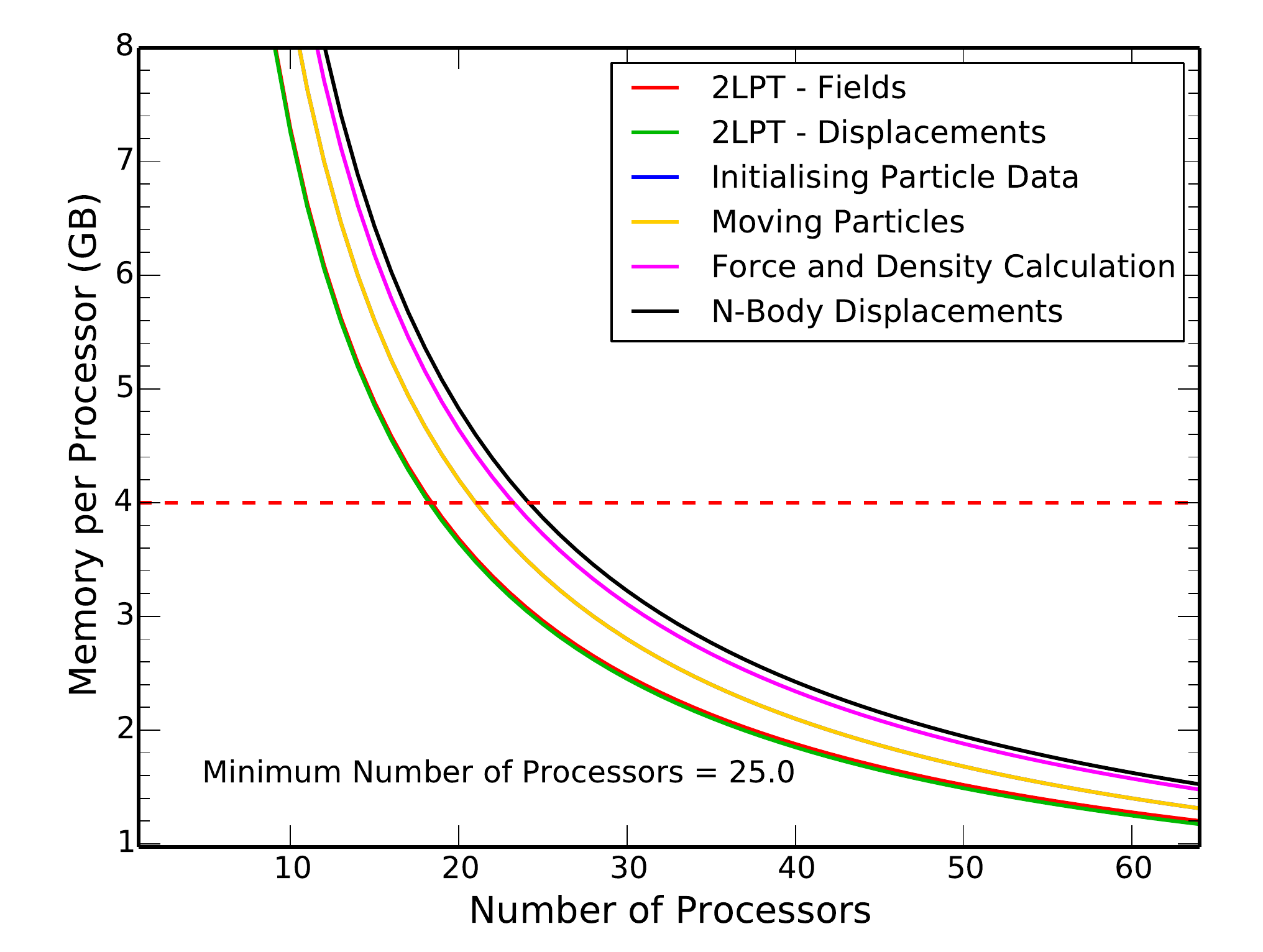}
  \caption{The memory requirements for the {\sc l-picola} run detailed in Section \ref{sec:accuracy}. The solid lines show the contributions from different sections of the code for varying numbers of processors, whilst the intersection of the dashed line with the solid lines gives the minimum number of processors required to run the simulation assuming there is 4GB of memory available per processor.}
  \label{picola_mem}
\end{figure}

As an example Figure \ref{picola_mem} shows the memory requirements as a function of number of processors for the {\sc l-picola} simulations used in Section \ref{sec:accuracy}. We can see that if we have 4GB of memory available per processor, this simulation can be run using only 25 processors if the optimal compilation options are used (32 were used for the simulations in this paper).


\begin{thebibliography}{99}

\bibitem[Ahn et al.(2012)]{Ahn2012} Ahn, C.~P., Alexandroff, 
R., Allende Prieto, C., et al.\ 2012, ApJS, 203, 21

\bibitem[Ahn et al.(2014)]{Ahn2014} Ahn, C.~P., Alexandroff, 
R., Allende Prieto, C., et al.\ 2014, ApJS, 211, 17 

\bibitem[Anderson et al.(2014)]{Anderson2014} Anderson, L., Aubourg, 
{\'E}., Bailey, S., et al.\ 2014, MNRAS, 441, 24 

\bibitem[Angulo et al.(2014)]{Angulo2014} Angulo, R.~E., Baugh, 
C.~M., Frenk, C.~S., \& Lacey, C.~G.\ 2014, MNRAS, 442, 3256 

\bibitem[Berlind 
\& Weinberg(2002)]{Berlind2002} Berlind, A.~A., \& Weinberg, D.~H.\ 2002, ApJ, 575, 587 

\bibitem[Bouchet et 
al.(1995)]{Bouchet1995} Bouchet, F.~R., Colombi, S., Hivon, E., \& Juszkiewicz, R.\ 1995, A\&A, 296, 575 

\bibitem[Chuang et al.(2015)]{Chuang2015} Chuang, C.-H., Kitaura, 
F.-S., Prada, F., Zhao, C., \& Yepes, G.\ 2015, MNRAS, 446, 2621 

\bibitem[Cole(1997)]{Cole1997} Cole, S.\ 1997, MNRAS, 286, 38 

\bibitem[Cole et al.(2005)]{Cole2005} Cole, S., Percival, W.~J., 
Peacock, J.~A., et al.\ 2005, MNRAS, 362, 505 

\bibitem[Coles 
\& Jones(1991)]{Coles1991} Coles, P., \& Jones, B.\ 1991, MNRAS, 248, 1

\bibitem[Colless et al.(2001)]{Colless2001} Colless, M., Dalton, 
G., Maddox, S., et al.\ 2001, MNRAS, 328, 1039 

\bibitem[Colless et al.(2003)]{Colless2003} Colless, M., Peterson, 
B.~A., Jackson, C., et al.\ 2003, arXiv:astro-ph/0306581 

\bibitem[Davis et al.(1985)]{Davis1985} Davis, M., Efstathiou, 
G., Frenk, C.~S., \& White, S.~D.~M.\ 1985, ApJ, 292, 371 

\bibitem[Dawson et al.(2013)]{Dawson2013} Dawson, K.~S., Schlegel, 
D.~J., Ahn, C.~P., et al.\ 2013, AJ, 145, 10

\bibitem[The Dark Energy Survey Collaboration(2005)]{DES2005} 
The Dark Energy Survey Collaboration 2005, arXiv:astro-ph/0510346

\bibitem[de la Torre 
\& Peacock(2013)]{delaTorre2013} de la Torre, S., \& Peacock, J.~A.\ 2013, MNRAS, 435, 743

\bibitem[Dodelson 
\& Schneider(2013)]{Dodelson2013} Dodelson, S., \& Schneider, M.~D.\ 2013, Phys. Rev. D, 88, 063537 

\bibitem[Drinkwater et al.(2010)]{Drinkwater2010} Drinkwater, M.~J., 
Jurek, R.~J., Blake, C., et al.\ 2010, MNRAS, 401, 1429 

\bibitem[Eisenstein et al.(2005)]{Eisenstein2005} Eisenstein, D.~J., 
Zehavi, I., Hogg, D.~W., et al.\ 2005, ApJ, 633, 560

\bibitem[Eisenstein et al.(2011)]{Eisenstein2011} Eisenstein, D.~J., 
Weinberg, D.~H., Agol, E., et al.\ 2011, AJ, 142, 72  

\bibitem[Feldman et al.(1994)]{Feldman1994} Feldman, H.~A., Kaiser, 
N., \& Peacock, J.~A.\ 1994, ApJ, 426, 23

\bibitem[Fosalba et al.(2013)]{Fosalba2013} Fosalba, P., Crocce, 
M., Gaztanaga, E., \& Castander, F.~J.\ 2013, arXiv:1312.1707  

\bibitem[Hockney \& Eastwood(1988)]{Hockney1988} Hockney, R. W., Eastwood, J. W., \textit{Computer Simulation Using Particles}, 1988, Adam Hilger

\bibitem[Howlett et al.(2012)]{Howlett2012} Howlett, C., Lewis, A., 
Hall, A., \& Challinor, A.\ 2012, J. Cosmo. Astroparticle Phys., 4, 027 

\bibitem[Howlett et al.(2015)]{Howlett2015} Howlett, C., Ross, 
A.~J., Samushia, L., Percival, W.~J., \& Manera, M.\ 2015, MNRAS, 449, 848 

\bibitem[Ivezic et al.(2008)]{Ivezic2008} Ivezic, Z., Tyson, 
J.~A., Abel, B., et al.\ 2008, arXiv:0805.2366

\bibitem[Jones et al.(2004)]{Jones2004} Jones, D.~H., Saunders, 
W., Colless, M., et al.\ 2004, MNRAS, 355, 747

\bibitem[Jones et al.(2009)]{Jones2009} Jones, D.~H., Read, 
M.~A., Saunders, W., et al.\ 2009, MNRAS, 399, 683 

\bibitem[Kaiser(1987)]{Kaiser1987}
  Kaiser N., 1987, MNRAS, 227, 1 

\bibitem[Kitaura 
\& He{\ss}(2013)]{Kitaura2013} Kitaura, F.-S., \& He{\ss}, S.\ 2013, MNRAS, 435, L78

\bibitem[Kitaura et al.(2014)]{Kitaura2014} Kitaura, F.-S., Yepes, 
G., \& Prada, F.\ 2014, MNRAS, 439, L21 

\bibitem[Klypin 
\& Holtzman(1997)]{1997astro.ph.12217K} Klypin, A., \& Holtzman, J.\ 1997, arXiv:astro-ph/9712217 

\bibitem[Laureijs et al.(2011)]{Laureijs2011} Laureijs, R., Amiaux, 
J., Arduini, S., et al.\ 2011, arXiv:1110.3193 

\bibitem[Lewis et al.(2000)]{Lewis2000} Lewis, A., Challinor, A., 
\& Lasenby, A.\ 2000, ApJ, 538, 473

\bibitem[Li et al.(2014)]{Li2014} Li, Y., Hu, W., 
\& Takada, M.\ 2014, Phys. Rev. D, 89, 083519  

\bibitem[Manera et al.(2013)]{Manera2013} Manera, M., Scoccimarro, 
R., Percival, W.~J., et al.\ 2013, MNRAS, 428, 1036 

\bibitem[Manera et al.(2015)]{Manera2015} Manera, M., Samushia, 
L., Tojeiro, R., et al.\ 2015, MNRAS, 447, 437 

\bibitem[Matsubara(1995)]{Matsubara1995} Matsubara, T.\ 1995, 
Progress of Theoretical Physics, 94, 1151 

\bibitem[Merson et al.(2013)]{Merson2013} Merson, A.~I., Baugh, 
C.~M., Helly, J.~C., et al.\ 2013, MNRAS, 429, 556 

\bibitem[Monaco et al.(2002)]{Monaco2002} Monaco, P., Theuns, T., 
Taffoni, G., et al.\ 2002, ApJ, 564, 8

\bibitem[Monaco et al.(2013)]{Monaco2013} Monaco, P., Sefusatti, 
E., Borgani, S., et al.\ 2013, MNRAS, 433, 2389 

\bibitem[Moutarde et al.(1991)]{Moutarde1991} Moutarde, F., Alimi, 
J.-M., Bouchet, F.~R., Pellat, R., \& Ramani, A.\ 1991, ApJ, 382, 377 

\bibitem[Pacheco(1997)]{Pacheco1997} Pacheco, P. S., \textit{Parallel Programming with MPI}, 1997, Morgan Kaufmann.

\bibitem[Peebles et al.(1989)]{Peebles1989} Peebles, P.~J.~E., 
Melott, A.~L., Holmes, M.~R., \& Jiang, L.~R.\ 1989, ApJ, 345, 108 

\bibitem[Percival et al.(2014)]{Percival2014} Percival, W.~J., Ross, 
A.~J., S{\'a}nchez, A.~G., et al.\ 2014, MNRAS, 439, 2531

\bibitem[Quinn et al.(1997)]{Quinn1997} Quinn, T., Katz, N., 
Stadel, J., \& Lake, G.\ 1997, arXiv:astro-ph/9710043 

\bibitem[Ross et al.(2015)]{Ross2015} Ross, A.~J., Samushia, L., 
Howlett, C., et al.\ 2015, MNRAS, 449, 835

\bibitem[Scoccimarro(1998)]{Scoccimarro1998} Scoccimarro, R.\ 1998, 
MNRAS, 299, 1097 

\bibitem[Scoccimarro 
\& Sheth(2002)]{Scoccimarro2002} Scoccimarro, R., \& Sheth, R.~K.\ 2002, MNRAS, 329, 629

\bibitem[Scoccimarro et al.(2012)]{Scoccimarro2012} Scoccimarro, R., 
Hui, L., Manera, M., \& Chan, K.~C.\ 2012, Phys. Rev. D , 85, 083002 

\bibitem[Seo \& Eisenstein(2003)]{Seo2003}
  Seo H.-J., Eisenstein D.J., 2003, ApJ, 598, 720 
  
\bibitem[Splinter et al.(1998)]{Splinter1998} Splinter, R.~J., 
Melott, A.~L., Shandarin, S.~F., \& Suto, Y.\ 1998, ApJ, 497, 38   
  
\bibitem[Springel(2005)]{Springel2005} Springel, V.\ 2005, MNRAS, 
364, 1105 

\bibitem[Takada 
\& Hu(2013)]{Takada2013} Takada, M., \& Hu, W.\ 2013, Phys. Rev. D, 87, 123504 

\bibitem[Tassev et al.(2013)]{Tassev2013} Tassev, S., Zaldarriaga, 
M., \& Eisenstein, D.~J.\ 2013, J. Cosmo. Astroparticle Phys., 6, 36

\bibitem[Tassev et al.(2015)]{Tassev2015} Tassev, S., Eisenstein, 
D.~J., Wandelt, B.~D., \& Zaldarriaga, M.\ 2015, arXiv:1502.07751

\bibitem[Taylor et al.(2013)]{Taylor2013} Taylor, A., Joachimi, 
B., \& Kitching, T.\ 2013, MNRAS, 432, 1928 

\bibitem[Tegmark(1997)]{Tegmark1997} Tegmark, M.\ 1997, Physical 
Review Letters, 79, 3806  

\bibitem[Tormen 
\& Bertschinger(1996)]{Tormen1996} Tormen, G., \& Bertschinger, E.\ 1996, ApJ, 472, 14 

\bibitem[White et al.(2014)]{White2014} White, M., Tinker, J.~L., 
\& McBride, C.~K.\ 2014, MNRAS, 437, 2594

\bibitem[Xu et al.(2012)]{Xu2012} Xu, X., Padmanabhan, N., 
Eisenstein, D.~J., Mehta, K.~T., \& Cuesta, A.~J.\ 2012, MNRAS, 427, 2146

\bibitem[York et al.(2000)]{York2000} York, D.~G., Adelman, J., 
Anderson, J.~E., Jr., et al.\ 2000, AJ, 120, 1579 

\bibitem[Zel'dovich(1970)]{Zeldovich1970} Zel'dovich, Y.~B.\ 1970, A\&A, 5, 84 

\end{thebibliography}
\end{document}